\documentclass{article}

\usepackage{arxiv}

\usepackage[utf8]{inputenc} %
\usepackage[T1]{fontenc}    %
\usepackage[hidelinks]{hyperref}       %
\usepackage{url}            %
\usepackage{booktabs}       %
\usepackage{amsfonts}       %
\usepackage{nicefrac}       %
\usepackage{microtype}      %
\usepackage{lipsum}
\usepackage{graphicx}

\usepackage{amsmath,mathrsfs,enumerate,multirow,appendix}
\usepackage{longtable, caption}
\usepackage{xcolor, url}
\usepackage{natbib}
\usepackage{bm}
\usepackage{placeins}

\usepackage[]{graphicx}
\chardef\bslash=`\\ %

\usepackage{tikz}

\usepackage{tikz-cd}
\tikzset{smalltext/.style={"\textup{\small #1}" description}}

\usetikzlibrary{shapes}
\usetikzlibrary{backgrounds}      %
\tikzcdset{
  my CD/.style={                  %
    arrow style=tikz,
    >={Triangle[length=2mm]},
    cells={nodes={inner sep=2mm}},
    row sep=1.5cm,  column sep=0.8cm}}
\tikzset{
  boxed/.style={                  
    show background rectangle,    %
    background rectangle/.append style={ %
      draw=gray, thick, rounded corners}}}

\makeatletter
\renewcommand\@biblabel[1]{#1.}
\makeatother

\graphicspath{ {./images/} }

\title{Correlation Matters! Streamlining the Sample Size Procedure with Hierarchical Time-to-event Endpoints}

\author{
 Yunhan Mou \\
  Department of Biostatistics,\\
  Yale School of Public Health\\
  New Haven, Connecticut, USA \\
  \texttt{yunhan.mou@yale.edu} \\
   \And
 Fan Li\\
    Department of Biostatistics,\\
    Center for Methods in Implementation and Prevention Science,\\
    Yale School of Public Health\\
    New Haven, Connecticut, USA \\
    \texttt{fan.f.li@yale.edu}\\
  \And
  Denise Esserman\\
    Department of Biostatistics,\\
    Yale Center for Analytical Sciences,\\
    Yale School of Public Health\\
    New Haven, Connecticut, USA\\
    \texttt{denise.esserman@yale.edu}\\
  \And
  Yuan Huang\thanks{Corresponding author}\\
    Department of Biostatistics,\\
    Yale Center for Analytical Sciences,\\
    Yale School of Public Health\\
    New Haven, Connecticut, USA\\
    \texttt{yuan.huang@yale.edu}\\
}

\usepackage{amsthm,amssymb,mathtools,float}
\usepackage[figuresright]{rotating}

\begin{document}
\maketitle
\vspace{-0.1in}
\begin{abstract}
{Composite endpoints are widely used in cardiovascular clinical trials to improve statistical efficiency while preserving clinical relevance. The win ratio (WR) measure and more general frameworks of win statistics have emerged as popular alternatives to traditional time-to-first-event analyses. Although analytic sample size formulas for WR have been developed, they rely on design parameters that are often not straightforward to specify. Consequently, sample size determination in clinical trials with WR for the primary analysis is most often based on simulations, which, however, can be computationally intensive. To address this challenge, we derive refined formulas to calculate the proportions of wins, losses, and ties for multiple prioritized time-to-event endpoints to assist with study planning. These formulas rely on familiar design inputs and become directly applicable when integrated with existing sample size methods. We conduct a comprehensive assessment of how correlation among endpoints affects sample size requirements across varying design features, and demonstrate the role of correlations through case studies to generate further insights.} %
\end{abstract}

\section{Introduction} \label{intro}
\subsection{Background and Literature Review}
Composite time-to-event endpoints are commonly employed in cardiovascular clinical trials to enhance clinical relevance and increase statistical efficiency. Popular examples include a combination of mortality and hospitalization, or the 3-point major adverse cardiovascular events (MACE) \citep{bosco2021major}, which consists of cardiovascular death, myocardial infarction, and stroke events. As an alternative to the traditional time-to-first-event analysis, which ignores the unequal clinical importance of involved endpoints, the win ratio (WR) measure \citep{pocock2012win} and the more general frameworks of win statistics (WS) \citep{dong2023win, verbeeck2023generalized, buyse2025handbook} (also referred to as the Generalized Pairwise Comparisons) have gained substantial attention in recent years; they rely on patient-level pairwise comparisons conducted hierarchically across endpoints of interest, with higher-priority outcomes (e.g., death) evaluated before lower-priority ones (e.g., hospitalization). Given its advantages, the WS framework has been recognized as an emerging approach for analyzing composite endpoints \citep{gasparyan2021adjusted}. For example, several clinical trials, including EMPULSE (NCT04157751), ATTRibute-CM (NCT03860935), VIP-ACS (NCT04001504), and ACTIVE (NCT03016975), have prespecified the WR measure for their primary outcome analysis. 
As the WR has gained traction in practical applications, methodological research has expanded in several directions including: methods for constructing confidence intervals and conducting hypothesis testing for the WR measure \citep{luo2015alternative, bebu2016large, dong2016generalized, mao2019alternative}; methods to accommodate stratification factors \citep{dong2018stratified, gasparyan2021adjusted}; and the statistical properties of the broader WS framework \citep{deltuvaite-thomas2022operational, buyse2025handbook}.

When employing the WR measure for primary analysis, determining an appropriate sample size is essential to ensure adequate statistical power while avoiding under- and over-enrolling participants. 
Several analytic sample size formulas for the WR measure have been proposed. \citet{yu2022sample} derived a closed-form expression by approximating the variance of the log-WR under a tie-group assumption (hereafter referred to as the YG formula).
{
\citet{mao2022sample} developed a more theoretically accurate formula based on U-statistics, introducing the standard rank deviation, which captures how much individual probabilities of winning vary across patients, as a key design parameter and providing simulation-based procedures for its estimation.
} \citet{zhou2022calculating} considered power estimation under the special case of two prioritized endpoints, requiring the solution of a polynomial equation, which can then be adapted for sample size calculation.
In addition, more discussion on the clinical trial design and further extensions to sample size calculation with other WS measures are available \citep{backer2024design, shu2024sample, barnhart2025trial, barnhart2025sample, lee2025notea, buyse2025handbook}.
{
For example, \citet{barnhart2025trial} showed that the overall WR can be expressed as a weighted average of the marginal (endpoint-specific) win ratios and derived closed-form expressions for the WR and probability of ties under the assumptions of independent exponential marginals, facilitating the use of the YG formula. \citet{barnhart2025sample} further extended the YG formula to accommodate the win odds, net benefit, and desirability of outcome ranking measures.} Nevertheless, the current practice in clinical trial design with WR as the primary measure remains simulation-based. In such approaches, datasets are repeatedly generated under prespecified design parameters, such as hazard ratios for time-to-event endpoints, and the required sample size is determined through back-calculation to achieve the desired power \citep{redfors2020win}. One key reason for this reliance on simulations is that the design parameters required by the available WR formulas 
{(e.g., tie probabilities in the YG formula or standard rank deviation in \citet{mao2022sample})}
are not easy to obtain a priori. 
{
Moreover, while between-endpoint correlation is often not fully described in published trial designs and may remain comparatively underexplored, its potential impact on the operating characteristics of the WR warrants careful consideration at the design stage.
}
For example, \citet{tang2024finkelsteina} demonstrated through simulations that, under the Finkelstein–Schoenfeld test, correlations among component endpoints can lead to markedly different study power. Using simulations with two time-to-event endpoints, \citet{mou2026generalizing} demonstrated that correlation between death and hospitalization can induce ``spurious'' negative treatment effects at the hospitalization outcome, arising from conditioning on ties at the death outcome, which leads to decreased power.
Other studies have also discussed how correlation affects the overall WR and the probability of ties across various combinations of endpoint types \citep{deltuvaite-thomas2022operational, fuyama2023impact, fuyama2025impacta}.
{
Importantly, correlations among component endpoints can be nontrivial in clinical trials, suggesting the necessity of accounting for them in study planning. For instance, \citet{mccarthy2020hometime} reported moderate correlations of home-time (i.e., time spent alive outside hospitals) with mortality (estimated Kendall's ${\tau} = 0.54$) and MACE (estimated Kendall's ${\tau} = 0.52$) in patients with type 2 myocardial infarction. In a more extreme case, \citet{greene2018hometime} found that post-discharge home-time was highly correlated with other time-to-event outcomes, with Kendall's ${\tau}$ reaching up to 0.8 for time to death and 0.6 for time to death or heart failure readmission.
}

\subsection{Motivations and Objectives}

Modern cardiovascular clinical trials increasingly prespecify hierarchical time-to-event endpoints, motivated by the need to capture the full patient event profile while preserving clinical priority. In practice, such endpoints are often selected because individual component events are relatively infrequent, and stakeholders value a treatment that meaningfully improves patient-centered hierarchies of benefit. 
For example, landmark trials like SPRINT \citep{thesprintresearchgroup2015randomized} and STICH \citep{velazquez2011coronaryartery} incorporated cardiovascular outcomes of varying clinical importance, including cardiovascular mortality, major non-fatal events, and hospitalizations, as primary or secondary endpoints.
{
Rather than relying on a traditional time-to-first-event analysis or focusing solely on a single primary endpoint, these outcomes can be analyzed using a prioritized structure, which is better aligned with clinical relevance and regulatory expectations.
In these studies, had win-based measures with hierarchy been the primary analysis, the investigators would have needed to account for the potential influence of correlation among time-to-event components, despite limited empirical guidance on how correlation impacts power and required sample size.
}

These considerations highlight a fundamental tension in trial design: while hierarchical time-to-event endpoints have clear interpretability advantages, difficulties in translating clinical intuition into formal sample size calculations remain, including the need for formula inputs that are not directly available and the frequent overlooking of correlation among endpoints.
To support data-driven and scalable design, there is a need for practical tools that make the impact of correlation transparent at the design stage.
{
To address this gap, we first propose a streamlined, formula-based procedure for sample size and power calculation for clinical trials with multiple hierarchical time-to-event endpoints, employing the YG formula. 
The procedure computes the probabilities of wins, losses, and ties using integral-based expressions, based on common clinical design specifications, such as hazard ratios, accrual periods, and dropout patterns, as well as additional specification of a copula to accommodate potential correlation among multiple prioritized time-to-event endpoints, and then incorporates these quantities into the YG formula for sample size and power estimation.
While the YG formula itself remains unchanged for sample size and power approximation in this procedure, the streamlined procedure allows more convenient specification of design assumptions, as the required inputs are more familiar to clinical investigators.
Second, we leverage the proposed procedure to investigate how correlation among endpoints influences required sample sizes under a range of design configurations. Specifically, we evaluate the trends under varying magnitudes of treatment effects across endpoints, different follow-up durations, and other practical design factors. While no simple universal rule emerges, our results indicate that the influence of correlation on power can be either favorable or unfavorable, with varying magnitudes, depending on the scenarios considered. This assessment provides useful initial insights into the potential impact of correlation on trial efficiency and power with multiple time-to-event endpoints.
}
Third, to facilitate practical use of our formulas, we develop an R package \texttt{SampleSizeWS} that implements the proposed formulas, power, and sample size estimation procedures, which allows for convenient real-world application (URL will be provided).  

\section{A Streamlined Procedure for Sample Size and Power Calculation}\label{method}

We consider a two-arm clinical trial with $N$ participants and $K$ prioritized time-to-event endpoints, using $t$ to denote treatment and $c$ to denote control. 
{
Let $Y_k^a$ denote the latent (censoring-free) event time for the $k$-th endpoint under arm $a\in\{t,c\}$, where endpoints are ordered by clinical priority ($k=1,2,\ldots,K$). We use $\mathcal{O}_k^a = \{\min(Y_k^a, C^a),\, I(Y_k^a \leq C^a)\}$ to denote the observed time and event status indicator for the $k$-th endpoint, where $C^a$ is the censoring time, and $\bm{\mathcal{O}}^a = (\mathcal{O}_1^a, \ldots, \mathcal{O}_K^a)$ to denote the collection of observed information from all $K$ endpoints.
For notational convenience, we let $S_a(\cdot)$ represent the joint, multivariate survival function of latent event times of all prioritized endpoints in arm $a$, and $h_a^{\partial k}(\cdot)$ denote the corresponding subdensity function for the $k$-th event such that $h_a^{\partial k}(x_1,x_2,\ldots,x_K) = -\frac{\partial}{\partial x_k} S_a(x_1,x_2,\ldots,x_K)$.
}
To compare two participants, we use $\succ$ to indicate better performance (a win), $\prec$ for worse performance (a loss), and $\simeq$ for a tie. Without loss of generality, we assume a longer time to an event is a more favorable outcome. 

\subsection{Review of the YG Formula}
We begin by briefly summarizing the YG formula that our procedure relies on. 
The YG formula uses the hypothesized overall $\text{WR}$ and the proportion of ties $p_{\text{tie}}$ as design parameters and offers companion formulas for estimating the required sample size with a given power or the resulting power for a fixed sample size \citep{yu2022sample}.
Among the $N$ participants, let $\rho\in(0,1)$ be the treatment allocation ratio. It holds that $\log(\widehat{\text{WR}})$ is asymptotically normal. Since the exact asymptotic variance is 
not straightforward for sample size planning, \cite{yu2022sample} derived a practical approximation given by
\begin{equation}
\mathrm{Var}\!\left[\log(\widehat{\text{WR}})\right]
\approx
\frac{\sigma^2}{N},
\quad
\sigma^2 = \frac{4(1+p_{\text{tie}})}{3\rho(1-\rho)(1-p_{\text{tie}})}.
\end{equation}
With the target effect size $\log({\text{WR}})$, the total sample size required for a two-sided test to achieve power $1-\beta$ at $\alpha$ level is then
\begin{equation}
N \approx \frac{\sigma^2\bigl(Z_{1-\alpha/2}+Z_{1-\beta}\bigr)^2}{\bigl[\log({\text{WR}})\bigr]^2},
\end{equation}
where $Z_{1-\alpha/2}$ and $Z_{1-\beta}$ are standard normal quantiles. 
Conversely, for a fixed $N$, the power is
\begin{equation} \label{eq:std_power}
    \text{Power} \approx 1 - \Phi\!\left( 
    Z_{1-\alpha/2} - \frac{|\log({\text{WR}})|\sqrt{N}}{\sigma} 
    \right) + \Phi\!\left( 
    -Z_{1-\alpha/2} - \frac{|\log({\text{WR}})|\sqrt{N}}{\sigma} 
    \right)
\end{equation}
with $\Phi(\cdot)$ the standard normal cumulative distribution function.

\subsection{Quantifying the Win, Loss and Tie Probabilities}
\label{sec:integration}
To compute the probabilities of wins, losses and ties, we consider independent censoring that may arise from multiple sources, such as administrative censoring and loss to follow-up, and assume the independent censoring times are identically distributed across arms.
Under this setup, the pairwise comparison between a treated and a control patient is made hierarchically across the prioritized endpoints. 
For the first endpoint $Y_1$, the treated individual is considered to have a \emph{win} when the event is observed in the control individual, and the event time occurs before any event or censoring in the treated individual, i.e.,
\[
Y_1^c < \min(Y_1^t, C^t, C^c),
\]
{
where $C^t$ and $C^c$ represent the identically  distributed  censoring times for the treated and control individuals, respectively.} Conversely, the treated individual is considered to have a \emph{loss} when the event is observed in the treated individual, and the event time occurs before any event or censoring in the control individual, i.e.,
\[
Y_1^t < \min(Y_1^c, C^c, C^t).
\]
A \emph{tie} arises when neither subject experiences the event before being censored, or when one is censored before the other’s event is observed, i.e.,
\[
C^t < \min(Y_1^t,Y_1^c,C^c)
\enspace \text{or} \enspace
C^c < \min(Y_1^t,Y_1^c,C^t).
\]
If the comparison on the first endpoint results in a tie, then the evaluation proceeds to the second outcome $Y_2$, and so forth, until either a win/loss is determined or all outcomes end in a tie. 

The win probability at the first endpoint can be expressed as:
\begin{equation}
\pi_{Y_1}^t 
    = \mathbb{P}(\mathcal{O}_1^t \succ \mathcal{O}_1^c) 
    = \mathbb{P}(Y_1^c < \min(Y_1^t, C^t, C^c))
    = \int_0^\infty S_t(y,0,\dots,0) S_{\widetilde{G}}(y)^2 
        h_c^{\partial 1}(y,0,\dots,0) dy,
\end{equation}
{
where $S_{\widetilde{G}}(\cdot)$ is the shared censoring survival function for $C^t$ and $C^c$.
Here, since all event times are non-negative, $S_t(y,0,\dots,0)=\mathbb{P}(Y_1^t>y, Y_2^t>0,\dots,Y_K^t>0)=\mathbb{P}(Y_1^t>y)$ reduces to the marginal survival function for the first endpoint in the treatment arm. }
An analogous expression holds for the loss probability $\pi_{Y_1}^c$, and the tie probability at the first endpoint is
\begin{align}
    \mathbb{P}(\mathcal{O}_1^t \simeq \mathcal{O}_1^c) &= 
        \mathbb{P}\left(C^t < \min(Y_1^t,Y_1^c,C^c)\right)
                        + \mathbb{P}(C^c < \min(Y_1^t,Y_1^c,C^t)), \notag \\
    &= 2\int_0^\infty S_t(c,0,\dots,0) S_c(c,0,\dots,0) \widetilde{g}(c) S_{\widetilde{G}}(c) dc,
\end{align}
{
where $\widetilde{g}(\cdot)$ is the shared probability density function of the censoring times.
}
For the $k$-th endpoint, with ties at the first through $(k-1)$-th endpoint, the win probability can be expressed as:
\begin{align}
    \pi_{Y_k}^t 
        &= \mathbb{P}\left(\mathcal{O}_k^t \succ \mathcal{O}_k^c, 
            \{\mathcal{O}_i^t \simeq \mathcal{O}_i^c\}_{i=1}^{k-1}\right) \notag\\ 
        &= \mathbb{P}\left(\mathcal{O}_k^t \succ \mathcal{O}_k^c, 
            C^t < \min(\{Y_i^t,Y_i^c\}_{i=1}^{k-1},C^c)\right) 
            + \mathbb{P}\left(\mathcal{O}_k^t \succ \mathcal{O}_k^c, 
            C^c < \min(\{Y_i^t,Y_i^c\}_{i=1}^{k-1},C^t)\right) \notag\\
        &= \mathbb{P}\left(Y_k^c < \min(Y_k^t, C^t, C^c), 
                C^t < \min(\{Y_i^t,Y_i^c\}_{i=1}^{k-1},C^c)\right) \notag\\
            &\quad+ \mathbb{P}\left(Y_k^c < \min(Y_k^t, C^t, C^c), 
                C^c < \min(\{Y_i^t,Y_i^c\}_{i=1}^{k-1},C^t)\right) \notag\\
        &= 2\int_{0}^{\infty} \widetilde{g}(c)   S_{\widetilde{G}}(c) \int_{0}^{c} 
                    S_t(\underbrace{c,\ldots,c}_{k-1}, y, 0,\ldots,0)
                    h_{c}^{\partial k}(\underbrace{c,\ldots,c}_{k-1},y,0,\dots,0) dy dc. 
\end{align}
Therefore, inputs required by the YG formula, namely the overall WR and the probability of ties, are obtained as:
\begin{eqnarray}
    &\text{WR} &= \frac{\sum_{k=1}^{K} \pi_{Y_k}^t}{\sum_{k=1}^{K} \pi_{Y_k}^c}, \\
    &p_{\text{tie}} &=\mathbb{P}(\bm{\mathcal{O}}^t \simeq \bm{\mathcal{O}}^c) = 
        2\int_0^\infty S_t(c,\dots,c)S_c(c,\dots,c)\widetilde{g}(c)S_{\widetilde{G}}(c)dc.
\end{eqnarray}
As a special case, when administrative censoring at a fixed follow-up time $s$ is the sole censoring mechanism, the general formulas reduce to a simpler yet slightly different form. The detailed expressions are provided in Supporting Information~S1.
{
In addition to the multiple time-to-event endpoint setting considered here, we further extend the proposed formulas to accommodate settings with multiple time-to-event endpoints followed by an ordinal endpoint; the corresponding details are provided in Supporting Information~S2.
}

{
In this setup, semi-competing risks, such as death censoring subsequent non-fatal endpoints, are not explicitly incorporated into the above formulas. To explicitly account for semi-competing risks, one may incorporate the death event as an additional censoring mechanism when computing the win, loss, and tie probabilities for subsequent non-fatal endpoints. In practice, the death event is typically placed on the first hierarchical layer due to its clinical importance and severity, which inherently mitigates the influence of semi-competing risks on the calculation of the proportions of wins, losses, and ties. Indeed, in the numerical studies in Section~3, we consider scenarios where the death event occupies the first hierarchical layer, and the proposed formulas continue to demonstrate satisfactory performance. 
A related but distinct issue arises when a competing risk is present, such as death from non-cardiovascular causes when the first layer is cardiovascular death. Under the standard WR procedure, a participant experiencing the competing event is treated as censored on all hierarchical layers, and hence the competing event enters our formulas as an additional component of the general censoring $\widetilde{G}$ under the same independent censoring assumption; this is the approach adopted in the case studies in Section~4.
}

{We make a final note that the restricted-time WR \citep{mao2024defining, wang2025restricted} provides a well-defined causal estimand by requiring all participants to be followed to a common restriction time, at least under the full-data scenario where latent event times are available. Our study centers on the standard WR approach, following the procedure of \citet{pocock2012win} and consistent with the sample-size proposals in \citet{yu2022sample} and \citet{barnhart2025sample}. In the presence of censoring, including
non-informative censoring, the standard WR estimator may be a biased estimator of the underlying restricted-time WR estimand \citep{mao2024defining}, as it instead targets a censoring-dependent estimand. One approach to achieve consistent estimation is adjustment via inverse probability of censoring weighting \citep{dong2020inverse}. Nevertheless, under equal censoring distributions across arms, both procedures coincide at the null, so the standard WR procedure remains a valid test of the strong null of no treatment effect. Since our study focuses on sample size and power calculation at the design stage with the assumption of independent censoring, we continue to use the standard WR estimation procedure throughout. This is also consistent with current practice of using WR in clinical trial design.}

\subsection{Design Specifications}
To make our framework more concrete, we show how $S_{\widetilde G}$ is constructed from typical censoring mechanisms in a clinical trial with a fixed study length $s$. 
For example, suppose participants are accrued over $[0,b]$ according to a Beta
distribution pattern, with entry time $A = b\,U$ where $U \sim \mathrm{Beta}(\psi_{\text{early}}, \psi_{\text{late}})$ and $\psi_{\text{early}}, \psi_{\text{late}}$ control the intensity of late versus early recruitment, respectively. In the absence of loss to follow-up, a participant entering at time $A$ is then followed for $L = s - A \sim s - b\,\mathrm{Beta}(\psi_{\text{early}},\psi_{\text{late}})$, whose
survival function is
\begin{equation}
    S_L(x) =
    \begin{cases}
    1, & x \le s - b, \\[6pt]
    F_{\mathrm{Beta}}\!\left( \dfrac{s-x}{b}; \psi_{\text{early}},\psi_{\text{late}} \right), & s - b < x \le s, \\[10pt]
    0, & x > s,
    \end{cases}
\end{equation}
where $F_{\mathrm{Beta}}\!\left( \cdot; \psi_{\text{early}},\psi_{\text{late}} \right)$ is the cumulative distribution function of the $\text{Beta}(\psi_{\text{early}},\psi_{\text{late}})$ distribution. As a special case, the uniform accrual period assumption, i.e., $\psi_{\text{early}}=\psi_{\text{late}}=1$, leads to:
\begin{equation}
    S_L(x) =
    \begin{cases}
        1, & x \le s-b, \\[6pt]
        \displaystyle \frac{s-x}{b}, & s-b < x \le s, \\[6pt]
        0, & x > s.
    \end{cases}
\end{equation}
Let the independent dropout censoring be represented by $G$ with survival function $S_G(\cdot)$ and probability density function $g(\cdot)$. Combined together, the general censoring $\widetilde G$ has the survival function $S_{\widetilde G}(x) = S_G(x) S_L(x)$ and probability density function $\widetilde g(x) = g(x)S_L(x) + f_L(x)S_G(x)$, where $f_L(\cdot)$ is the probability density function that can be obtained directly from $S_L(\cdot)$.

To design a clinical trial, the joint survival functions for prioritized time-to-event endpoints under each treatment arm, $S_t$ and $S_c$, are also needed. These can be constructed by first specifying the marginal distribution of each endpoint under common clinical design assumptions, most often exponential or piecewise exponential distributions, parameterized by hazard rates, event rates, or hazard ratios, and then combining the marginals through a copula to incorporate correlation across endpoints. Formally, let $S_{a,k}(y) = \mathbb{P}(Y_k^a > y)$ denote the marginal survival function for endpoint $k$ under arm $a \in \{t,c\}$. Common specifications may include the exponential model:
\begin{equation} \label{exp_margin}
    S_{a,k}(y) = \exp(-\lambda_{a,k} y),
\end{equation}
or, a piecewise exponential model with breakpoints $0=\eta_0 < \eta_1 < \cdots < \eta_J=+\infty$,  
\begin{equation}
S_{a,k}(y) = \exp\left(- \sum_{j=1}^{J} \lambda_{a,k,j} \big(\min(y,\eta_j) - \eta_{j-1}\big)_+\right),
\end{equation}
where $\lambda_{a,k},\lambda_{a,k,j}$ are hazard rates, and $(\cdot)_+$ is the positive part function. 
For even greater flexibility, the marginal survival function $S_{a,k}(y)$ may also be specified in a nonparametric or semiparametric manner, such as by employing estimated Kaplan–Meier curves or model-based smooth estimators using preliminary data.

To account for the correlation across endpoints, we then link the marginals through a copula structure. For example, employing the Gumbel--Hougaard copula \citep{nelsen2006introduction}, we have:
\begin{equation}\label{GH_copula}
    S_a(y_1,\ldots,y_K) 
    = \exp\!\left\{-\Bigg[ \sum_{k=1}^K \big(-\log S_{a,k}(y_k)\big)^{\kappa} \Bigg]^{1/\kappa}\right\}, 
    \quad \kappa \ge 1,
\end{equation}
leading to the Kendall’s $\tau = 1 - 1/\kappa$.
This Gumbel-Hougaard copula is straightforward to specify and is often adequate when the goal is to control the overall correlation level with positive dependence. However, if a more detailed specification of pairwise correlations across endpoints is required, or if negative dependence is anticipated, the Gaussian copula provides a more flexible alternative:
\begin{equation} \label{Gaussian_copula}
S_a(y_1,\ldots,y_K) 
= \Phi_{\boldsymbol{\Sigma}}\!\left(\Phi^{-1}(S_{a,1}(y_1)), \ldots, \Phi^{-1}(S_{a,K}(y_K))\right),
\end{equation}
where $\Phi^{-1}(\cdot)$ is the standard normal quantile function and $\Phi_{\boldsymbol{\Sigma}}$ is the cumulative distribution function of a mean zero multivariate normal distribution with correlation matrix $\boldsymbol{\Sigma}$. The elements of the correlation matrix control the Kendall's $\tau$. 

\subsection{Extension to Stratified Designs}
We further extend our sample size procedure to stratified designs. Assume that the participants are divided into $M$ strata (e.g., by study site, region, or baseline covariates), and pairwise comparisons are conducted within each stratum.
For the $m$-th stratum, the probabilities of wins and losses at each endpoint, as well as the overall probability of ties, are calculated as described in Section~\ref{sec:integration}, and are denoted by $\pi_{Y_k}^{t,m}$, $\pi_{Y_k}^{c,m}$ ($k=1,2,\ldots,K$), and $p_{\text{tie}}^m$, respectively.
Let $N_{m}$ be the total number of participants in the $m$-th stratum, $\rho$ be the shared allocation ratio, and $w_{m}$ be the stratum-specific weight.
The overall WR and probability of ties across strata, denoted by $\text{WR}_{\text{strata}}$ and $p_{\text{tie}}^{\text{strata}}$, are then calculated as:
\begin{eqnarray}
    &\text{WR}_{\text{strata}} &= 
        \frac{\sum_{m=1}^M w_m\left[
            \left( \sum_{k=1}^K \pi_{Y_k}^{t,m}\right) N_m^2 
            \right]}
            {\sum_{m=1}^M w_m\left[
                \left( \sum_{k=1}^K \pi_{Y_k}^{c,m}\right) N_m^2 
                \right]}, \\
    &p_{\text{tie}}^{\text{strata}} &=
        \frac{\sum_{m=1}^M w_m\, p_{\text{tie}}^m\, N_m^2}
            {\sum_{m=1}^M w_m N_m^2}.
\end{eqnarray}
The stratified YG formula approximates the variance of $\log(\widehat{\text{WR}}_{\text{strata}})$ as:
\begin{equation}
\mathrm{Var}\!\left[\log(\widehat{\text{WR}}_{\text{strata}})\right]
\approx 
\frac{4(1+p_{\text{tie}}^{\text{strata}})}
    {3\rho(1-\rho)(1-p_{\text{tie}}^{\text{strata}})}
\cdot
\frac{\sum_{m=1}^M w_m^2 N_m^3}
    {\left(\sum_{m=1}^M w_m N_m^2\right)^2}.
\end{equation}
This leads to a power formula similar to equation~(\ref{eq:std_power}) and could be employed for sample size estimation under a stratified design.

\section{Numerical Studies}
We conduct numerical investigations to evaluate the streamlined sample size procedure and to explore the impact of correlation on study power. We begin by comparing simulation-based empirical results with the corresponding quantities obtained from the analytic formulas, including statistical power, the overall WR, and the probability of ties. We then leverage the formulas to systematically examine the influence of correlation among endpoints on the required sample size, exploring how variations in treatment effects and follow-up time affect statistical efficiency.

In our numerical studies, we adopt the following general setup. The Gumbel-Hougaard copula with exponential marginals is employed, i.e., equations (\ref{exp_margin}) and (\ref{GH_copula}). The hazard rates are constructed by $\lambda_{t,k}=\lambda_{c,k}\exp(-\alpha_k),k=1,2,\ldots,K$, where $\lambda_{c,k}$ are the baseline hazard rates in the control group per day and $\alpha_k$ controls the treatment effects (corresponding to a hazard ratio of $\exp(-\alpha_k)$ at the $k$-th endpoint). $\lambda_{c,k}$ and $\alpha_k$ are selected to ensure that the overall scaling reflects possible real-world clinical trial settings \citep{voors2022sglt2}. Dropout censoring is represented by an independent exponential distribution with a hazard of 0.00015 per day. The scheduled maximal follow-up time $s$ is varied from 250 to 1500 days, with a uniform accrual period of 200 days. The proposed formulas are calculated via numerical integration with R programming. A 1:1 allocation ratio and a two-sided significance level of 0.05 are used throughout. The detailed setup of each scenario is presented subsequently.

\subsection{Evaluation of the Streamlined Sample Size Procedure}
We begin by evaluating two aspects of the proposed streamlined procedure. The first focuses on the accuracy of the formula-based estimates of the overall WR and probability of ties; the second examines the performance of our power calculator. For both, the results obtained from the streamlined procedure are compared with those from the conventional simulation-based approach. Under the streamlined procedure, the assumed design parameters are directly used to calculate the probabilities of winning, losing, and tying, and subsequently the power through the YG formula. In contrast, the simulation-based method begins with the same design assumptions but generates observed datasets repeatedly under these settings. For each replicate, pairwise comparisons are conducted to obtain the empirical probabilities of winning, losing, and tying. The simulation-based probabilities are then computed as the averages across all replicates, and the empirical power is determined as the proportion of simulations rejecting the null hypothesis.

We consider three prioritized time-to-event endpoints, with baseline hazard rates (per day) in the control arm set to 
$
\bm\lambda_c = \left(\lambda_{c,1},\lambda_{c,2},\lambda_{c,3} \right) 
                = \left(0.00057, 0.0018, 0.0015\right).
$ 
The total sample size is fixed at $N=1000$. To investigate the impact of correlations across endpoints, we vary the overall Kendall’s $\tau$, with $\tau \in \{0,0.3,0.5,0.8\}$. Simulation-based results are obtained from 2000 replicates under each scenario with R programming and package \texttt{WinRatio} \citep{package_WinRatio}. Here, we consider three configurations of treatment effect magnitudes {for the non-null scenarios}, parameterized by
$
\bm \alpha = \left(\alpha_1,\alpha_2,\alpha_3 \right) \in \{ (0.2,0.3,0.1), (0.1,0.2,0.3)
$, $(0.2,0.2,0.2)\}$,
representing two different patterns of varying effects across the prioritized endpoints alongside a balanced scenario. {The null scenario with $\boldsymbol{\alpha} = (0,0,0)$ is also considered. }

{
For the ease of presentation,} Table~\ref{tab:SIM1} compares the overall WR and the probability of ties 
obtained from the integral-based formulas 
{ with their simulation-based estimates
for selected combinations of Kendall's $\tau (\tau \in \{ 0, 0.3, 0.8\} )$ and maximal follow-up time $s (s \in \{ 500, 1000\} )$ under the non-null scenarios. Results  from  the remaining scenarios are provided in Supporting Information~S3 and show patterns consistent with those reported here.} Across all scenarios, the two approaches yield nearly identical results, confirming the accuracy of our formulas in reproducing empirical quantities. It is worth noting that in the simulation, we consider settings where the top event censors the occurrence of the remaining two events, whereas this feature is not explicitly addressed in the integral-based formulation. The close agreement nevertheless highlights the broad applicability and robustness of the proposed formulas.

\begin{table}[h!]
  \centering
  \caption{Estimated $\text{WR}$ and $p_{\text{tie}}$ obtained from formulas ($\widetilde{\text{WR}},\widetilde{p_{\text{tie}}}$) and simulations 
  ($\widehat{\text{WR}},\widehat{p_{\text{tie}}}$) across treatment effect magnitudes ($\bm \alpha$),
  correlation levels (in Kendall's $\tau$), and maximal follow-up times ($s$).}
    \begin{tabular}{lrrrrrr}
    \toprule
    $\bm \alpha$ & \multicolumn{1}{l}{$s$} & \multicolumn{1}{l}{$\tau$} & \multicolumn{1}{l}{$\widetilde{\text{WR}}$} & \multicolumn{1}{l}{$\widehat{\text{WR}}$} & \multicolumn{1}{l}{$\widetilde{p_{\text{tie}}}$(\%)} & \multicolumn{1}{l}{$\widehat{p_{\text{tie}}}$(\%)} \\
    \midrule
    (0.2, 0.3, 0.1) & 500   & 0     & 1.26  & 1.27  & 11.00 & 10.92 \\
    (0.2, 0.3, 0.1) & 500   & 0.3   & 1.26  & 1.26  & 18.24 & 18.09 \\
    (0.2, 0.3, 0.1) & 500   & 0.8   & 1.28  & 1.29  & 31.38 & 31.22 \\
    (0.2, 0.3, 0.1) & 1000  & 0     & 1.26  & 1.27  & 4.27  & 4.25 \\
    (0.2, 0.3, 0.1) & 1000  & 0.3   & 1.27  & 1.27  & 6.22  & 6.20 \\
    (0.2, 0.3, 0.1) & 1000  & 0.8   & 1.29  & 1.29  & 11.50 & 11.47 \\
    (0.1, 0.2, 0.3) & 500   & 0     & 1.20   & 1.20   & 11.17 & 11.10 \\
    (0.1, 0.2, 0.3) & 500   & 0.3   & 1.21  & 1.21  & 18.50 & 18.34 \\
    (0.1, 0.2, 0.3) & 500   & 0.8   & 1.22  & 1.22  & 31.44 & 31.28 \\
    (0.1, 0.2, 0.3) & 1000  & 0     & 1.16  & 1.16  & 4.31  & 4.30 \\
    (0.1, 0.2, 0.3) & 1000  & 0.3   & 1.17  & 1.17  & 6.30  & 6.25 \\
    (0.1, 0.2, 0.3) & 1000  & 0.8   & 1.19  & 1.19  & 11.53 & 11.50 \\
    (0.2, 0.2, 0.2) & 500   & 0     & 1.22  & 1.22  & 10.97 & 10.88 \\
    (0.2, 0.2, 0.2) & 500   & 0.3   & 1.22  & 1.23  & 18.17 & 18.00 \\
    (0.2, 0.2, 0.2) & 500   & 0.8   & 1.22  & 1.23  & 31.08 & 30.89 \\
    (0.2, 0.2, 0.2) & 1000  & 0     & 1.22  & 1.22  & 4.26  & 4.26 \\
    (0.2, 0.2, 0.2) & 1000  & 0.3   & 1.22  & 1.23  & 6.20  & 6.15 \\
    (0.2, 0.2, 0.2) & 1000  & 0.8   & 1.22  & 1.23  & 11.35 & 11.31 \\
    \bottomrule
    \end{tabular}%
  \label{tab:SIM1}%
\end{table}%

\begin{figure}[ht!]
\centerline{\includegraphics[width=6in]{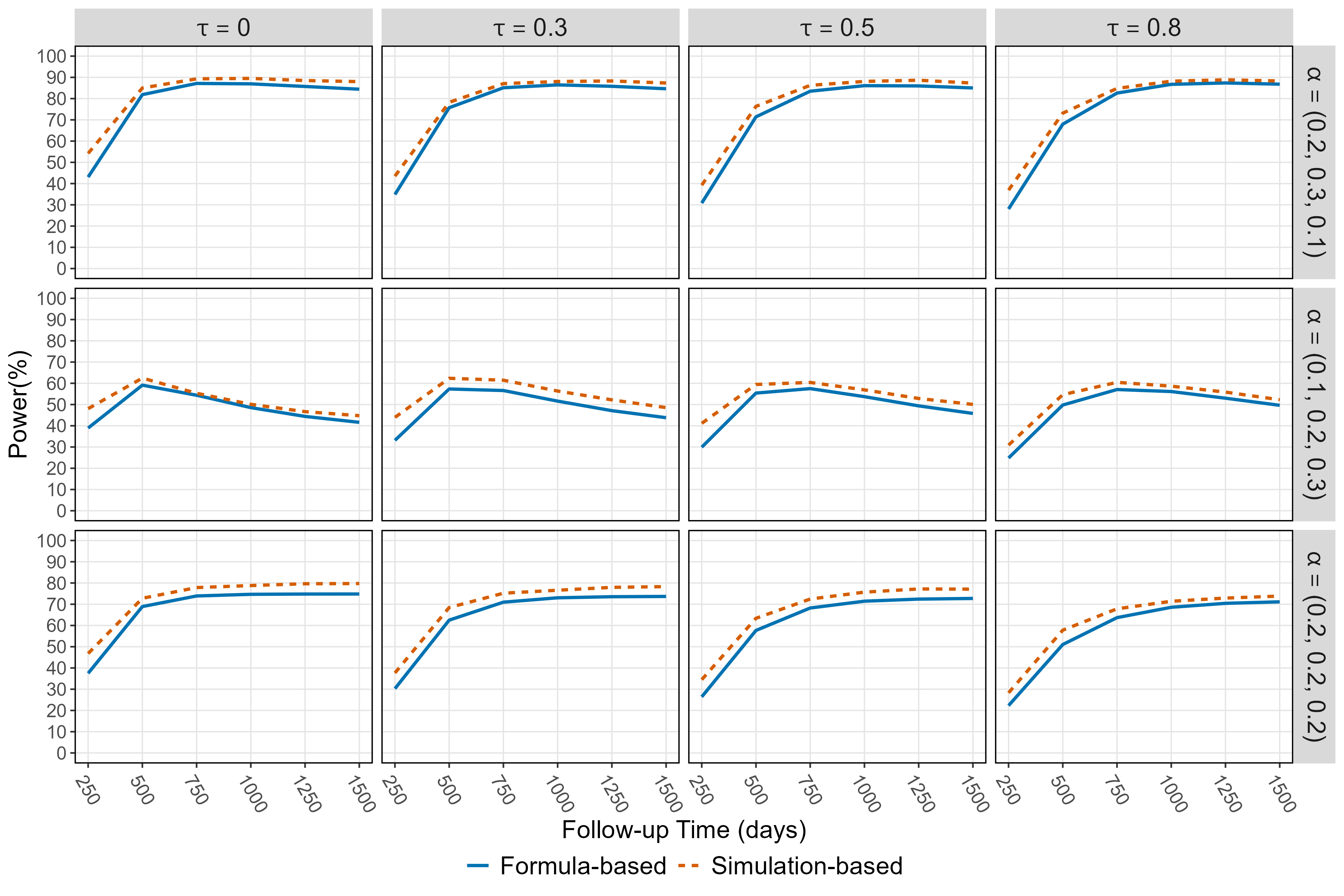}}
\caption{Formula-based and simulation-based estimation of power across treatment effect magnitudes ($\bm \alpha$), correlation levels ($\tau$), and follow-up times. \label{fig:SIM1}}
\end{figure}

Figure~\ref{fig:SIM1} displays the empirical power obtained from simulation and the corresponding estimates from the proposed formulas. 
{
Across all scenarios, the formula-based estimates closely track the simulation-based estimates, with the former being slightly conservative. This conservatism is more pronounced with a smaller $s$, that is, when $p_{\text{tie}}$ is larger; for example, at a follow-up time of 250 days, the empirical $p_{\text{tie}}$ ranges from 46.66\% to 67.81\% across scenarios.} 
{Since the overall WR and $p_{\text{tie}}$ obtained from the integral-based formulas and simulations are nearly identical, this conservative difference likely reflects a feature of the YG formula itself, consistent with the findings of \citet{lee2025notea} based on different, though related, simulation studies.} 
{
Additionally, when the treatment effect is stronger on the lower-priority endpoints, as in the scenario with $\bm \alpha=(0.1,0.2,0.3)$,  power first increases as follow-up is extended, but then gradually decreases with longer follow-up. This is because at the earlier stage, extending follow-up leads to more observed events across all three endpoints, thereby resolving more tied pairwise comparisons and thus improving power. At a later stage, however, the additional follow-up increasingly resolves ties at the top-priority endpoint, where the treatment effect is weaker. This reduces the relative contribution of the lower-priority endpoints, which carry stronger treatment effects, to the overall pairwise comparison results; this in turn attenuates the power.}

\subsection{Impact of Between-Endpoint Correlation for Study Planning}  \label{subsec:cor4planning}

We next investigate the influence of between-endpoint correlation on the required sample size based on the proposed streamlined procedure. Here, we consider two prioritized time-to-event endpoints, with baseline hazard rates (per day) in the control arm set to  
$
\bm\lambda_c = (0.00057,0.0015).
$ 
{
Treatment effects are specified by   
$
\bm \alpha \in \{ (0.3,0.1),\allowbreak (0.3,0.05),\allowbreak (0.2,0.2),\allowbreak (0.1,0.3),
$ and $(0.05,0.3)\}$, representing scenarios where treatment effect magnitudes are either stronger on the first endpoint, stronger on the second, or identical across both.} The maximal follow-up time is set to $s \in \{500,1000,1500\}$, and the positive correlation between endpoints is varied from $\tau =0$ to $0.9$ by an increment of $0.1$. Sample sizes are calculated to achieve 90\% and 80\% power, corresponding to type II error rates of 0.1 and 0.2, respectively.

\begin{figure}
\centerline{\includegraphics[width=6in]{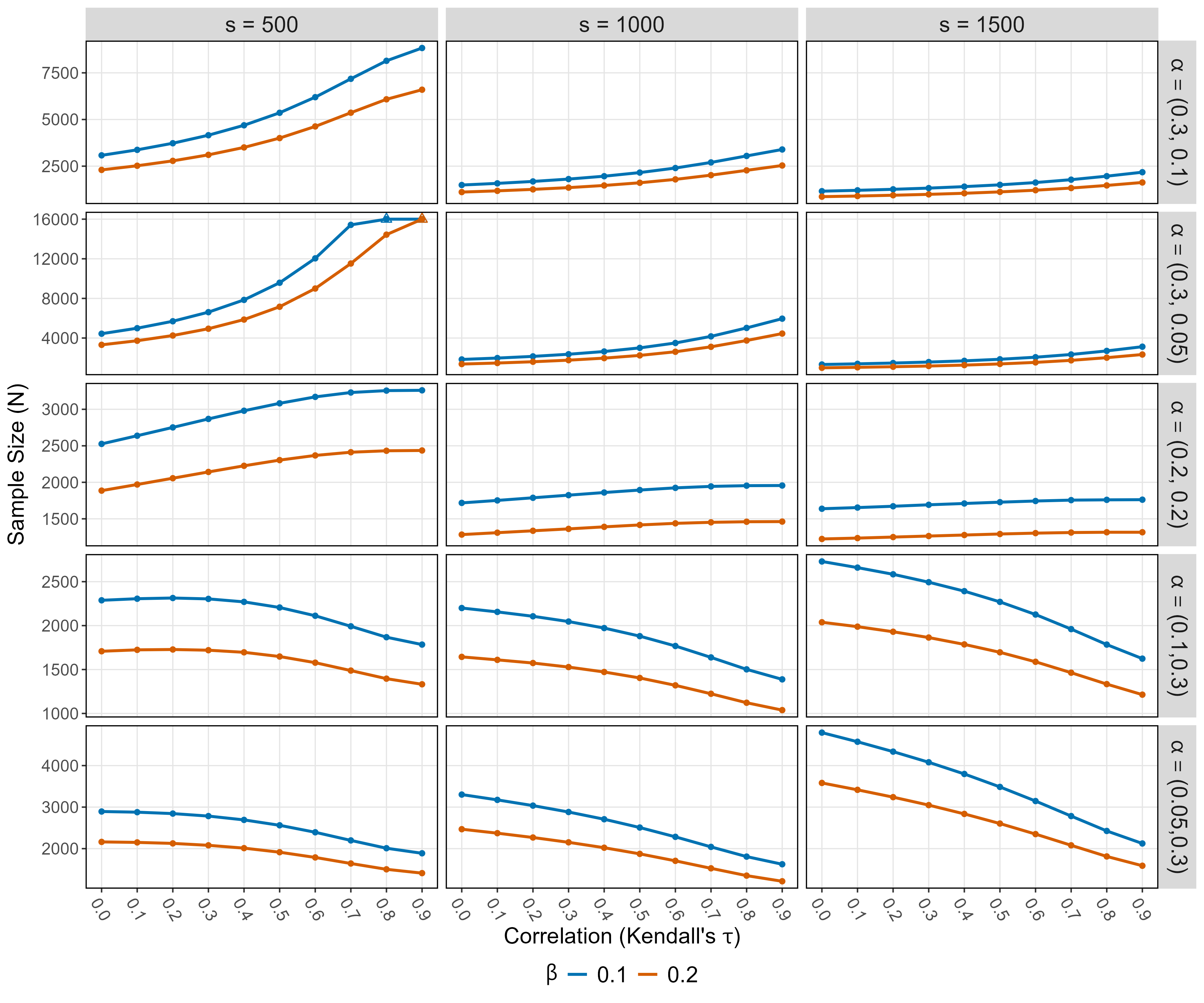}}
\caption{
{Formula-based sample size estimation across treatment effect magnitudes ($\bm \alpha$), correlation levels ($\tau$), follow-up times ($s$), and targeting type II error rates ($\beta$). For better visualization, extreme values of sample size exceeding 16,000 are marked by hollow triangles.} \label{fig:SIM2_SS}}
\end{figure}

{
Figure~\ref{fig:SIM2_SS} displays the estimated sample sizes across different correlation levels.  Several general trends are observed. First, when the treatment effect is stronger on the first endpoint, increasing correlation tends to require a larger sample size. In contrast, when the treatment effect is stronger on the second endpoint, higher correlation typically reduces the required sample size. This difference reflects the relative contribution of the correlated endpoints to the overall WR, depending on where the treatment effect is concentrated. Second, the impact of correlation may depend on the length of follow-up. When the first layer carries a stronger signal, longer follow-up may yield more observed events at that layer, thereby concentrating pairwise comparison outcomes within it. Hence correlation may have a reduced impact. Conversely, when the signal at the top layer is weak, longer follow-up may amplify the impact of correlation.  
}

\begin{sidewaysfigure}
\centerline{\includegraphics[width=\textwidth]{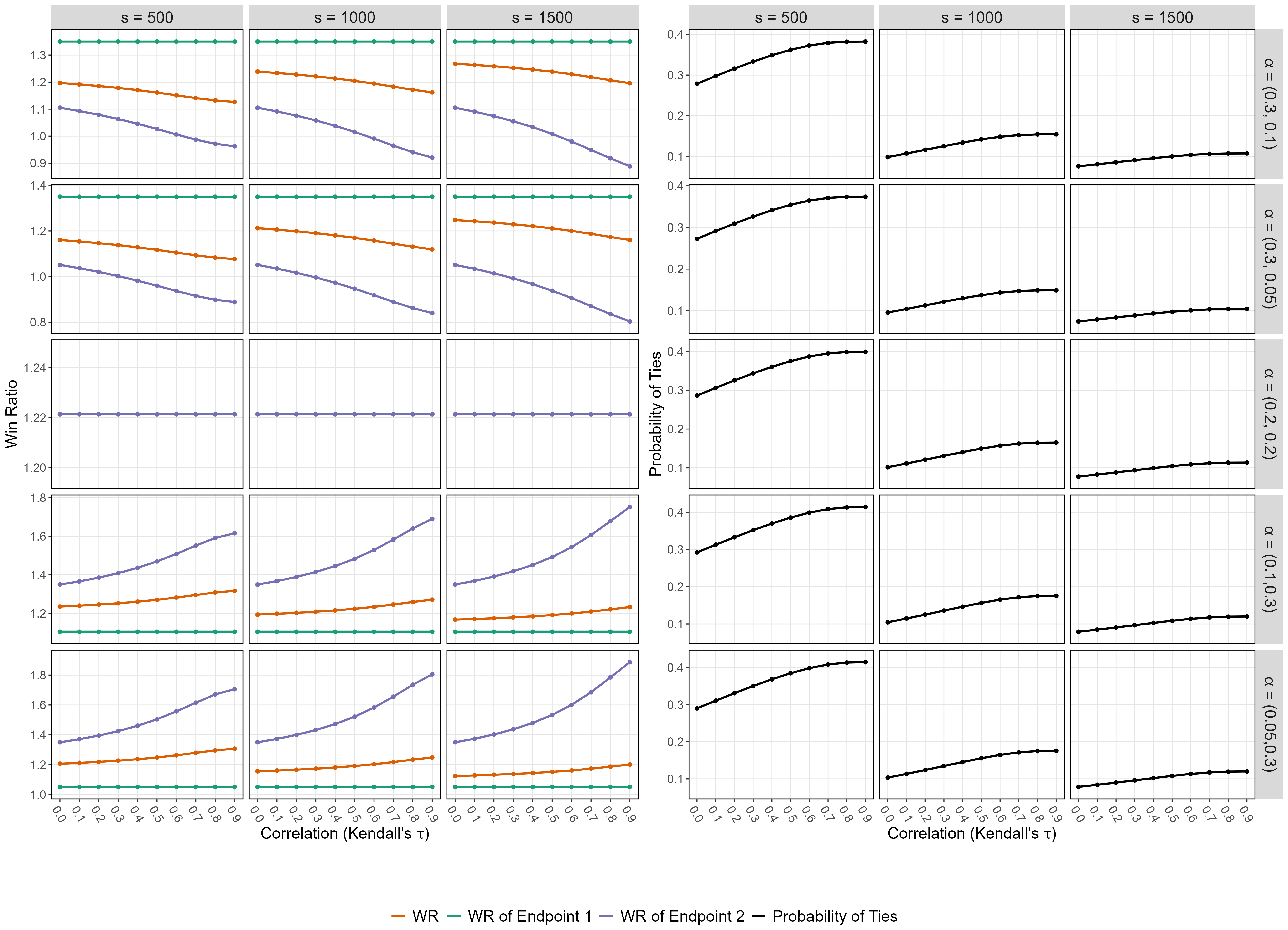}}
\caption{{Win ratio (WR) and probability of ties across simulation settings. \textit{Left panel:} overall and endpoint-specific WR.
\textit{Right panel:} probability of ties ($p_{\text{tie}}$).
Both are shown across treatment effect magnitudes ($\bm{\alpha}$),
correlation levels ($\tau$), and follow-up times ($s$).} \label{fig:SIM2_WR_ptie_trend}}
\end{sidewaysfigure}

Figure~\ref{fig:SIM2_WR_ptie_trend} summarizes the trends in WR and $p_{\text{tie}}$ across correlation levels.
For all scenarios, a higher correlation consistently leads to higher tie probabilities, as expected. The increase is more substantial for a shorter follow-up length $s$.  When treatment effects are stronger on the first endpoint, increasing correlation leads to decreased overall WR, driven largely by a reduction in the contribution of the second endpoint-specific WR. This reduction in the overall WR increases the sample size needed to attain the same level of power.
In contrast, when the treatment effect is stronger on the second endpoint, correlation has an opposite effect, which raises the second endpoint-specific WR and the overall WR, thereby reducing the required sample size.
{
When the treatment effect magnitudes are the same across endpoints, higher correlation continues to increase $p_{\text{tie}}$, but has no impact on the overall WR, which remains constant at 1.22 across all correlation levels.
}

To gain further insights into the mechanisms driving these trends, we next examine several representative scenarios in detail. 
First, in the case of $s=500$ and $\bm{\alpha}=(0.3,0.1)$, the required sample size increases gradually at low correlation levels and then more steeply as the correlation strengthens, with the rate of increase easing somewhat at the highest correlation levels. This pattern reflects the combined influence of the two underlying quantities: the decline in the overall WR accelerates as correlation increases, whereas the rise in $p_{\text{tie}}$ is most rapid at low correlation and gradually plateaus thereafter. The accelerating reduction in the overall WR therefore becomes the dominant driver of the sample size requirement at moderate to high correlation levels.
{Second, for the scenario with $s=1000$ and $\bm{\alpha}=(0.2,0.2)$, where the treatment effect magnitudes on the two endpoints are identical, correlation generally increases the required sample size. This pattern is primarily driven by the rise in $p_{\text{tie}}$, while the overall WR remains stable.
It is also notable that, in this scenario, the endpoint-specific WR and the overall WR are the same, and they remain unchanged across different follow-up times $s$.}
{
This is expected because, under the equal-signal configuration with the Gumbel--Hougaard copula and exponential margins, the two endpoints satisfy a joint proportional hazards structure with a common proportionality factor, under which the WR remains invariant across endpoints and follow-up durations \citep{oakes2016win}.
}
Third, when $s=500$ and $\bm{\alpha}=(0.1,0.3)$, the overall trend shows that increasing correlation lowers the required sample size. However, a slight increase in sample size is observed at low correlation levels ($\tau \leq 0.2$). In this setting, the overall WR increases with a higher correlation, which reduces the required sample size, but a higher correlation also increases $p_{\text{tie}}$, which acts in the opposite direction by inflating the sample size. When $\tau \leq 0.2$, the influence of $p_{\text{tie}}$ slightly outweighs that of the WR, leading to the initial rise. 

{Finally, in addition to the formula-based sample size estimation, we also perform simulation-based sample size estimation for the same set of scenarios with the fixed study-length setup, targeting $\beta=0.2$. The corresponding results are presented in Supporting Information~S4.1 and demonstrate patterns consistent with those observed under the formula-based approach. We further conduct numerical studies under a fixed follow-up time setup with varying levels of dropout censoring, and the corresponding results are provided in Supporting Information~S4.2. The results exhibit patterns similar to those observed under the fixed study-length setup, with the fixed follow-up time setup further confirming that heavier dropout censoring leads to larger required sample sizes through an elevated $p_{\text{tie}}$.}

\section{Case Studies}

While the numerical studies in Section 3 establish the operating characteristics of the proposed framework under controlled conditions, practical trial design requires grounding in real clinical settings. 
{
To this end, we examine two landmark cardiovascular studies, SPRINT and STICH, where multiple time-to-event endpoints (either primary or secondary) are available and could potentially be analyzed or structured under a win-based design framework.
}
Specifically, we illustrate the application of the proposed streamlined procedure and showcase the impact of between-endpoint correlation in a real-world context under a two-endpoint setup for both studies. 
These two trials differ in their design features and thus serve as complementary representative cases, exhibiting distinct patterns of correlation influence on the required sample size. They also span different tie-probability settings: SPRINT is characterized by a higher overall $p_{\text{tie}}$, whereas STICH exhibits a lower overall $p_{\text{tie}}$, such that the two cases also illustrate the behavior of the YG formula across different tie-probability regimes.
Supporting Information~S5.2 further considers a three-endpoint extension based on the STICH trial, with the dependence structure specified using a Gaussian copula. 

In the design phase, the specifications of design parameters often need to rely on pilot data, clinical relevance, or assumptions elicited from stakeholders. However, as the case analyses presented are solely for the purposes of demonstration, we consider the design with no stratification and estimate the sample size in a post hoc manner by calculating 
key design parameters from the trial itself.
More specifically, the hazard rates of each time-to-event endpoint for the treatment and control groups are estimated separately under the exponential distribution assumption.
{
Using the data from both treatment and control groups, we then estimate the correlation among component endpoints in Kendall's $\tau$ with a parametric approach assuming the Gumbel structure, as described in Section~2.5 of \citet{emura2010goodnessoffit} for right-censored endpoints, implemented via the \texttt{Copula.surv} R package (version 2.0) using the \texttt{U2.Gumbel} function.
}
The hazard rate for dropout is similarly estimated under the exponential assumption using the trial data. The scheduled maximal follow-up time is set to the longest observed follow-up, and the accrual period is adopted from the corresponding trial publication {with a uniform entry assumption}. A two-sided test with a significance level of 0.05 is considered throughout.

\subsection{The SPRINT Trial}
The Systolic Blood Pressure Intervention Trial (SPRINT) \citep{thesprintresearchgroup2015randomized} was a large, randomized controlled trial designed to evaluate whether intensive systolic blood pressure (SBP) control, which targeted SBP below 120 mm Hg, reduces cardiovascular morbidity and mortality compared with a standard SBP target below 140 mm Hg in adults without diabetes.
A total of 9,361 participants at increased cardiovascular risk were randomized to intensive or standard treatment and followed for a median of 3.26 years. The primary composite endpoint was the time to first occurrence of myocardial infarction (MI), other acute coronary syndromes (ACS), stroke, heart failure (HF), or cardiovascular-related death.
In the original analysis, the intensive treatment group experienced a significantly lower rate of the primary composite outcome. 
In our analysis, we consider a hierarchical endpoint with two layers: (1) time to cardiovascular-related death as the first layer, and (2) time to first non-fatal cardiovascular event (MI, ACS, stroke, or HF) as the second layer. This setup reflects a conventional hierarchy that assigns higher priority to fatal events over non-fatal events.
The endpoint-level proportions of wins, losses, and ties for the included endpoints in the observed SPRINT data are presented in Supporting Information~S5.1.

Based on the SPRINT dataset and publication \citep{thesprintresearchgroup2015randomized}, the accrual period was approximately 881 days, and the maximum follow-up time was 1,744 days. The estimated hazard rate of dropout (including non-cardiovascular death) was $6.71 \times 10^{-5}$ per day. 
For the two prioritized endpoints, the estimated hazard rates in the treatment group were $6.73 \times 10^{-6}$ and $4.02 \times 10^{-5}$ per day, respectively, while the corresponding hazard rates in the control group were $1.19 \times 10^{-5}$ and $4.95 \times 10^{-5}$ per day. The correlation between the two endpoints, estimated using a Gumbel copula model, was $\hat{\tau} = 0.1886$.
The sample size calculations based on the estimated design parameters under varying correlation levels are presented in Table~\ref{tab:apx_SPRINT_SS}. For both type II error levels, the required sample sizes show a monotonic increase as the correlation between endpoints strengthens. For example, relative to the independence scenario, having the empirically estimated correlation ($\hat{\tau} = 0.1886$) from the SPRINT dataset requires approximately 23\% more participants to achieve the same target power. 
This pattern mirrors what was observed in the simulation scenarios where the treatment effect is stronger on the first endpoint (i.e., $\bm{\alpha} = (0.3,0.1)$ and $\bm{\alpha} = (0.3,0.05)$). In this case analysis, the estimated hazard ratios for the fatal and composite non-fatal endpoints are 0.57 and 0.81, respectively, indicating a stronger treatment effect on the fatal endpoint. Furthermore, both endpoints exhibit relatively low event rates, leading to an estimated overall $p_{\text{tie}}$ ranging from 89\% to 91\%. Such a high prevalence of tied pairs also contributes to the amplified influence of correlation on the required sample size.
When such a sample size table is presented at the design stage, it calls on investigators to recognize and account for the potential influence of correlation among endpoints. Referring to correlation levels reported in comparable studies or estimated from pilot data can help investigators better approximate the required sample sizes to achieve the desired power. 

{
Beyond the role of correlation, this case study also warrants attention to the magnitude of the sample size estimates themselves. The high overall tie probability places the YG formula in the regime where its conservatism becomes particularly pronounced.
This conservatism is directly shown by the comparison between the formula- and simulation-based sample size estimates at $\beta = 0.2$ in Table~\ref{tab:apx_SPRINT_SS}, where the simulation-based estimates are obtained following the same procedure as that described in Supporting Information~S4.1.
The formula-based estimates exceed their simulation-based counterparts by roughly 35\% to 47\% across the correlation levels considered. Importantly, however, even under such pronounced conservatism of the YG formula, the proposed procedure continues to capture the influence of correlation on the required sample size correctly. This is shown through the relative sample size, defined as the required sample size at each correlation level divided by that obtained under the independence assumption ($\tau = 0$), which isolates the trend attributable to correlation from the absolute scale of the estimates. As reported in Table~\ref{tab:apx_SPRINT_SS}, the formula- and simulation-based approaches yield close relative sample sizes across the entire range of $\tau$, indicating that the conservatism of the YG formula affects the magnitude but not the shape of the correlation-driven trend.
When $p_{\text{tie}}$ is anticipated to be high at the design stage, as in the present case, we recommend first employing the streamlined procedure to characterize the trend of how correlation influences the required sample size, and then obtaining simulation-based sample size estimates. In our experience, using the formula-based sample sizes as starting values for the simulation search can also substantially accelerate this procedure, since the conservative formula-based estimates provide an 
informative upper bound for the target sample size and thereby help narrow the search range.
}

\begin{table}[htbp]
  \centering
  \caption{
    Formula- and simulation- based sample size estimates for the SPRINT trial under varying correlation levels ($\tau$) and type II error rates ($\beta$). The estimates corresponding to the independence assumption ($\tau = 0$) and to the empirically estimated correlation level ($\hat{\tau} = 0.1886$) are highlighted in bold. The relative sample size is the required sample size at each $\tau$ divided by that under $\tau = 0$.
  }
  \resizebox{\textwidth}{!}{%
    \begin{tabular}{lrrrrrrrrrrr}
    \toprule
          & \multicolumn{11}{c}{$\tau$} \\
          & \textbf{0} & 0.1   & \textbf{0.1886} & 0.2   & 0.3   & 0.4   & 0.5   & 0.6   & 0.7   & 0.8   & 0.9 \\
    \midrule
    \multicolumn{12}{l}{\textit{Formula-based}} \\
    $\beta = 0.1$ & 13342 & 14850 & 16376 & 16586 & 18564 & 20754 & 23024 & 25078 & 26492 & 27064 & 27176 \\
    $\beta = 0.2$ & 9966  & 11092 & 12232 & 12390 & 13866 & 15502 & 17198 & 18734 & 19790 & 20216 & 20300 \\
    \multicolumn{12}{l}{\textit{Simulation-based}} \\
    $\beta = 0.2$ & 7010  & 7940  & 9030  & 9050  & 9810  & 10750 & 12400 & 12780 & 13450 & 14080 & 14480 \\
    \multicolumn{12}{l}{\textit{Relative sample size ($\beta = 0.2$, referenced to $\tau = 0$)}} \\
    Formula-based & 1.00  & 1.11  & 1.23  & 1.24  & 1.39  & 1.56  & 1.73  & 1.88  & 1.99  & 2.03  & 2.04 \\
    Simulation-based & 1.00  & 1.13  & 1.29  & 1.29  & 1.40  & 1.53  & 1.77  & 1.82  & 1.92  & 2.01  & 2.07 \\
    \bottomrule
    \end{tabular}%
  }
  \label{tab:apx_SPRINT_SS}%
\end{table}%

\subsection{The STICH Trial} \label{sec:STICH2}
The Surgical Treatment for Ischemic Heart Failure (STICH) trial \citep{velazquez2011coronaryartery} was a randomized controlled trial conducted across 22 countries, enrolling 1,212 patients with coronary artery disease and left ventricular systolic dysfunction (ejection fraction $\leq$ 35\%). 
Participants were assigned to receive either optimal medical therapy alone or medical therapy plus coronary-artery bypass grafting (CABG). The trial was designed to evaluate whether the addition of CABG to guideline-based medical therapy could reduce mortality in this high-risk population. The primary endpoint was all-cause mortality, while key secondary endpoints included death from cardiovascular causes and the composite of death from any cause or hospitalization for cardiovascular causes. Although the primary endpoint did not differ significantly between groups, adding CABG was associated with lower rates of cardiovascular death and of death or hospitalization for cardiovascular causes.
{
We consider a hierarchical endpoint with two layers: (1) time to cardiovascular-related death, followed by (2) time to first cardiovascular-related hospitalization.
The endpoint-level proportions of wins, losses, and ties for the included endpoints in the observed STICH data are presented in Supporting Information~S5.1.
}
In STICH, the accrual period was approximately 1,746 days, and the maximum follow-up time was 3,051 days. The estimated hazard rate of dropout (including non-cardiovascular death) was $3.88 \times 10^{-5}$ per day. 
For the two prioritized endpoints, the estimated hazard rates in the treatment group were $1.92 \times 10^{-4}$ and $3.22 \times 10^{-4}$ per day, respectively, while the corresponding hazard rates in the control group were $2.38 \times 10^{-4}$ and $5.11 \times 10^{-4}$ per day. The correlation between the two endpoints is estimated to be $\hat{\tau} = 0.1578$. 
The required sample sizes based on these design parameters under different correlations with the Gumbel--Hougaard copula are presented in Table~\ref{tab:STICH_SS}. For both type II error levels, we observe that the required sample sizes exhibit a modest non-monotonic pattern with respect to the correlation level. The sample size increases slightly at low to modest correlation levels and then gradually decreases as the correlation strengthens. This behavior represents a milder version of the pattern observed in the simulation scenario with $s=500$ and $\bm{\alpha}=(0.1,0.3)$. The trend can be attributed to the stronger treatment effect on the hospitalization endpoint, reflected by an estimated hazard ratio of 0.63, compared with 0.81 for the death endpoint, along with the relatively high event rates observed in this trial (leading to the estimated overall $p_{\text{tie}}$ ranging from {14\% to 26\%}). The contribution of the second layer becomes more prominent than the increased $p_{\text{tie}}$ at higher correlation levels, thereby leading to a gradual reduction in the required sample size. When such a sample size table is presented at the design stage, the independence
assumption then clearly emerges as a safe default, since the resulting sample sizes are comparable to or more conservative than those under the correlated case.
We further note that in this STICH case study, the formula-based sample size estimates are closely aligned with their simulation-based counterparts, which is likely attributable to the relatively low $p_{\text{tie}}$ in this setting, where the conservativeness of the YG formula is less pronounced than in the SPRINT case study. As expected, the two approaches yield the same pattern regarding the influence of correlation on the required sample size.

\begin{table}[htbp]
  \centering
  \caption{
    Formula- and simulation- based sample size estimates for the STICH trial under varying correlation levels ($\tau$) and type II error rates ($\beta$). The estimates corresponding to the independence assumption ($\tau = 0$) and to the empirically estimated correlation level ($\hat{\tau} = 0.1578$) are highlighted in bold. The relative sample size is the required sample size at each $\tau$ divided by that under $\tau = 0$.
  }
  \resizebox{\textwidth}{!}{%
    \begin{tabular}{lrrrrrrrrrrr}
    \toprule
          & \multicolumn{11}{c}{$\tau$} \\
          & \textbf{0} & 0.1   & \textbf{0.1578} & 0.2   & 0.3   & 0.4   & 0.5   & 0.6   & 0.7   & 0.8   & 0.9 \\
    \midrule
    \multicolumn{12}{l}{\textit{Formula-based}} \\
    $\beta = 0.1$ & 768   & 774   & 776   & 778   & 780   & 778   & 770   & 750   & 718   & 674   & 632 \\
    $\beta = 0.2$ & 574   & 578   & 580   & 582   & 584   & 582   & 576   & 562   & 538   & 504   & 472 \\
    \multicolumn{12}{l}{\textit{Simulation-based}} \\
    $\beta = 0.2$ & 510   & 510   & 530   & 520   & 510   & 520   & 500   & 500   & 480   & 460   & 420 \\
    \multicolumn{12}{l}{\textit{Relative sample size ($\beta = 0.2$, referenced to $\tau = 0$)}} \\
    Formula-based & 1.00  & 1.01  & 1.01  & 1.01  & 1.02  & 1.01  & 1.00  & 0.98  & 0.94  & 0.88  & 0.82 \\
    Simulation-based & 1.00  & 1.00  & 1.04  & 1.02  & 1.00  & 1.02  & 0.98  & 0.98  & 0.94  & 0.90  & 0.82 \\
    \bottomrule
    \end{tabular}%
  }
  \label{tab:STICH_SS}%
\end{table}%

\section{Discussion} \label{sec:discussion}

In this study, we propose a streamlined procedure for sample size and power calculation when multiple prioritized time-to-event endpoints are considered, with greater emphasis on between-endpoint correlations. We hereby provide several additional remarks and extensions. First, it is noteworthy that the YG formula generally yields power estimates comparable to those obtained from simulation-based methods, though it may be moderately conservative in certain scenarios. This pattern becomes more pronounced when the overall $p_{\text{tie}}$ is high, as observed in our numerical studies (e.g., the larger discrepancies observed at $s=250$). In practical trial design, while the combination of derived formulas and the YG formula provides direct and computationally efficient sample size estimation, it is advisable to perform a supplementary simulation-based power assessment using the estimated sample size as a validation step. This cross-checking can be helpful in studies with low event rates, where the analytic approach tends to be more conservative. Nonetheless, the streamlined procedure remains advantageous in such cases by narrowing the simulation effort to a targeted set of candidate sample sizes, rather than requiring an exhaustive, potentially computationally prohibitive grid search. 
{
On the other hand, even in regimes where the YG formula exhibits more pronounced conservatism, our experience suggests that the general trend of how correlation influences the required sample size (e.g., whether increasing correlation leads to a larger, stable, or smaller required sample size) remains adequately reflected by the YG formula. In this sense, the proposed procedure remains a valuable computationally efficient tool for understanding the role of correlation in study planning, even though the associated absolute sample size estimates warrant simulation-based confirmation in scenarios with high $p_{\text{tie}}$.
}

{Second, we employ copula structures to represent the between-endpoint correlation and thereby specify the joint survival distribution. It is possible that the choice of copula may influence the resulting sample size and power approximation, even when the marginal distributions are held fixed. For instance, the Gumbel--Hougaard copula, which arises from the positive stable frailty model, exhibits upper tail dependence and thus induces stronger dependence among early event times. In contrast, the Clayton--Oakes copula, which arises from the gamma frailty model, possesses lower tail dependence and therefore captures stronger dependence among late event times \citep{nelsen2006introduction}. As a result, given a finite follow-up period, the Gumbel--Hougaard copula may show higher influence of correlation on sample size requirements relative to the Clayton--Oakes copula, even when the marginals are identical. In contrast, the Gaussian copula imposes a normality assumption directly on transformed marginal distributions of the event times and exhibits symmetric, asymptotically vanishing tail dependence.
In practice, validating the copula assumption on the joint distribution at the design stage is inherently difficult, even when pilot data are available, as such data are typically limited in size for reliably assessing the joint dependence structure of several time-to-event endpoints. Throughout this article, we adopt the Gumbel--Hougaard and Gaussian copulas for illustration, both of which are convenient for sample size and power approximation. The proposed formulas, however, accommodate other copula specifications, and alternative structures may be preferable when clinical prior knowledge is available to inform the dependence pattern; for example, the Clayton--Oakes copula may be adopted when the event times are believed to exhibit stronger dependence at later stages.
}

{Third, our findings highlight the value of routinely reporting between-endpoint correlations in clinical trials employing WS. As we have shown throughout, the correlation among prioritized endpoints can materially shape the overall WR, the probability of ties, and ultimately the required sample size; yet, unlike marginal quantities such as hazard ratios or event rates, empirical estimates of between-endpoint association are seldom reported in current practice, leaving investigators with little to anchor their assumptions at the design stage. Sensitivity analyses across a plausible range of correlation values remain indispensable in the face of this uncertainty. Nevertheless, association estimates from previously completed trials with comparable endpoints and populations often constitute the most credible starting point when dedicated pilot data are unavailable. This situation is reminiscent of the trajectory in the literature for cluster randomized trials, where the central role of the intracluster correlation coefficient in sample size determination motivated explicit reporting recommendations and spurred the assembly of empirical correlation repositories to support future designs \citep{ouyang2023estimating}. To build a stronger evidence base for designing clinical trials with complex endpoints, we therefore embed a similar recommendation within the WS framework: broader and more systematic reporting of empirical between-endpoint correlations for composite time-to-event endpoints (e.g., through the natural association measures within the Gumbel--Hougaard and
Gaussian copula models) would improve the realism of design assumptions and, over time, deepen the collective understanding of how component outcomes co-vary in different contexts.}

{
Fourth, we acknowledge that the selection of endpoints is of vital importance for clinical trials employing WS analyses. Although the hierarchical endpoint structure provides a convenient and clinically grounded framework for incorporating multiple outcome measures in the primary analysis, this should not be treated as a substitute for careful endpoint selection. While it may appear that there is little cost to including an additional endpoint in the hierarchy, the trade-offs are more nuanced than they first appear. As our study clearly demonstrates, correlation among component endpoints can meaningfully influence the overall WR, probability of ties, and consequently the power, which represents one concern among others that arise when expanding the endpoint hierarchy. 
For example, if a candidate endpoint is known to be highly correlated with an endpoint already included in the hierarchical structure, one must carefully weigh the potential benefits of additional information against the possible negative consequences. 
More broadly, both clinical relevance and statistical considerations should be thoroughly evaluated at the endpoint selection stage, ensuring that each included endpoint contributes meaningful information to the overall assessment of treatment benefit.
}

{Finally, we comment on the generalizability of the proposed procedure beyond the setting illustrated here, along two dimensions: the types of endpoints and the choice of win statistic. On endpoint types, our main development concerns multiple prioritized time-to-event endpoints, and we additionally address a time-to-event hierarchy followed by an ordinal endpoint (Supporting Information~S2), a structure that arises frequently in cardiovascular trials when a recurrent-event count or ordinal measure is appended below the survival components. We acknowledge that clinical trials across different domain areas may nonetheless involve a broader range of endpoint types, including continuous, categorical, and longitudinal outcomes; developing corresponding formulas for the proportions of wins, losses, and ties under such mixed structures is an important direction for future work that would further broaden the applicability of the method. On the choice of win statistic, because the YG formula was originally developed for the WR, we have centered our illustration on the WR for concreteness. We emphasize, however, that the first-stage computation of the win, loss, and tie probabilities depends only on the hierarchical endpoints themselves and not on the treatment effect measure. Once these probabilities are obtained, other win statistics such as the net benefit \citep{buyse2010generalized} and the win odds \citep{dong2020winodds} follow directly, and can be combined with their corresponding sample size and power formulas (derived as transformations of the YG
formula \citep{barnhart2025sample}) to form analogous streamlined procedures. In this sense, the win, loss, and tie probabilities serve as a common, modular foundation from which a range of WS designs can be constructed.}

\section{Supplementary Material}
\label{sec6}

Supporting Information is included in the appendix of this manuscript.

\section*{Acknowledgments}
This work is partially supported by the United States National Institutes of Health (NIH), National Heart, Lung, and Blood Institute (NHLBI, grant number 1R01HL178513). All statements in this report, including its findings and conclusions, are solely those of the authors and do not necessarily represent the views of the NIH.
Yunhan Mou's research is supported by CTSA Grant Number UL1 TR001863 from the National Center for Advancing Translational Sciences (NCATS), a component of the National Institutes of Health (NIH). Its contents are solely the responsibility of the authors and do not necessarily represent the official view of NIH.
The authors thank the SPRINT and STICH study teams for making the data available through the Biologic Specimen and Data Repository Information Coordinating Center (BioLINCC) at the National Heart, Lung, and Blood Institute (NHLBI). 
{
The corresponding BioLINCC accession numbers are HLB01701818b for STICH and HLB02021925a for SPRINT.
}

\bibliographystyle{biom}
\bibliography{refs}

\clearpage
{\LARGE \textbf{Supporting Information}}
\appendix
\renewcommand{\thesection}{S\arabic{section}}
\renewcommand{\thesubsection}{S\arabic{section}.\arabic{subsection}}
\renewcommand{\thetable}{S\arabic{table}}
\renewcommand{\thefigure}{S\arabic{figure}}
\setcounter{table}{0}
\setcounter{figure}{0}
\renewcommand{\theHtable}{S\arabic{table}}
\renewcommand{\theHfigure}{S\arabic{figure}}
\section{Calculation of Win Loss and Tie Probabilities with Administrative Censoring}
\label{apx:admin_censor_only}

In this section, we consider the scenario where censoring arises solely from administrative censoring at a fixed follow-up time $s$. In this setting, each TTE endpoint is either observed prior to $s$ or censored at $s$. Pairwise comparisons between a treated and a control patient are made hierarchically across the prioritized endpoints. For the first outcome $\mathcal{O}_1$, the treatment participant scores a win if 
$Y_1^c < \min(Y_1^t, s)$,
a loss if 
$Y_1^t < \min(Y_1^c, s)$,
and a tie if both event times exceed $s$. If the comparison on the first outcome results in a tie, then the evaluation proceeds to the second outcome $\mathcal{O}_2$, and so forth, until either a win/loss is determined or all outcomes tie.  

Under this setup, the win probability of a treatment participant at the first endpoint can be expressed as  
\[
\pi_{Y_1}^t = \mathbb{P}\left(Y_1^c < \min(Y_1^t, s)\right) = 
    \mathbb{P}\left(Y_1^t > Y_1^c, Y_1^c <s\right) 
    = \int_0^s S_t(y,0,\dots,0) h_c^{\partial 1}(y,0,\dots,0) dy,
\]  
with an analogous expression for the loss probability $\pi_{Y_1}^c$. Tie probabilities arise naturally when both latent event times exceed the censoring time, i.e.,  
\[
\mathbb{P}(\mathcal{O}_1^t \simeq \mathcal{O}_1^c) = \mathbb{P}(Y_1^t > s, Y_1^c > s).
\]  
These definitions extend recursively to subsequent endpoints, where comparisons are conditioned on the tie status of earlier outcomes. The win probability at the $k$-th endpoint is then:
\begin{align*}
    \pi_{Y_k}^t &= \mathbb{P}(Y_k^t > Y_k^c, Y_k^c < s, 
                                \{Y_i^t > s, Y_i^c > s\}_{i=1}^{k-1})\\
        &= \int_0^s S_t(s,\dots,s,y,0,\dots,0) 
                    h_c^{\partial k}(s,\dots,s,y,0,\dots,0) dy.
\end{align*}

The overall WR and tie probability are then given by:
\begin{eqnarray*}
    &\text{WR} &= \frac{\sum_{k=1}^{K} \pi_{Y_k}^t}{\sum_{k=1}^{K} \pi_{Y_k}^c},\\
    &p_{\text{tie}} &= \mathbb{P}(\bm{\mathcal{O}}^t \simeq \bm{\mathcal{O}}^c) = S_t(s,\dots,s)S_c(s,\dots,s).
\end{eqnarray*}

\section{
{Calculation of Win Loss and Tie Probabilities with Time-to-event and Ordinal Endpoints}
}
\label{apx:TTE_cont_endpoint}

In this section, we extend the proposed procedure to settings with $K$ prioritized time-to-event endpoints and an ordinal outcome at the lowest-priority level, such as a clinical summary score or functional class assessed at the end of follow-up. Without loss of generality, we let the $(K+1)$-th endpoint, $Y_{K+1}$, take values in $\{1,2,\dots,Q\}$, where larger values indicate more favorable outcomes. For arm $a \in \{t,c\}$, we define the marginal survival function of the ordinal endpoint as $S_{a,K+1}(q)=\mathbb{P}(Y_{K+1}^a>q)$. By definition, $S_{a,K+1}(0)=1$ and $S_{a,K+1}(Q)=0$. The joint survival function for all $K+1$ endpoints may then be specified using the copula-based constructions described in the main text. To accommodate possible missingness in the ordinal endpoint, let $R_{K+1}\in\{0,1\}$ denote the observation indicator for $Y_{K+1}$, where $R_{K+1}=1$ indicates that the ordinal outcome is observed and $R_{K+1}=0$ indicates that it is missing. We assume a common observation mechanism across treatment arms, with $\mathbb{P}(R_{K+1}=1)=\xi$, and consider the scenario in which missingness is completely at random (MCAR). For the pairwise comparison on the $(K+1)$-th endpoint to be conclusive, both participants in the pair are required to have $R_{K+1}=1$; pairs in which at least one unit has $R_{K+1}=0$ are treated as ties on this endpoint.

For the first endpoint ($k=1$), the win and loss probabilities are:
\begin{align}
    \pi_{Y_1}^t 
        &= \int_{0}^{\infty} S_t(y, 0, \ldots, 0, 0) S_{\widetilde{G}}(y)^2 h_c^{\partial 1}(y, 0, \ldots, 0, 0) dy, \\
    \pi_{Y_1}^c 
        &= \int_{0}^{\infty} S_c(y, 0, \ldots, 0, 0) S_{\widetilde{G}}(y)^2 h_t^{\partial 1}(y, 0, \ldots, 0, 0) dy.
\end{align}
For the subsequent time-to-event endpoints ($k \in \{2, 3, \ldots, K\}$), the win and loss probabilities incorporate this marginalization as follows:
\begin{align}
    \pi_{Y_k}^t 
        &= \mathbb{P}\left(\mathcal{O}_k^t \succ \mathcal{O}_k^c, 
            \{\mathcal{O}_i^t \simeq \mathcal{O}_i^c\}_{i=1}^{k-1}\right) \notag\\ 
        &= 2\int_{0}^{\infty} \widetilde{g}(c)   S_{\widetilde{G}}(c) \int_{0}^{c} 
                    S_t(\underbrace{c,\ldots,c}_{k-1}, y, 0,\ldots,0, 0)
                    h_{c}^{\partial k}(\underbrace{c,\ldots,c}_{k-1},y,0,\dots,0, 0) dy dc, \\
    \pi_{Y_k}^c 
        &= \mathbb{P}\left(\mathcal{O}_k^t \prec \mathcal{O}_k^c, 
            \{\mathcal{O}_i^t \simeq \mathcal{O}_i^c\}_{i=1}^{k-1}\right) \notag\\ 
        &= 2\int_{0}^{\infty} \widetilde{g}(c)   S_{\widetilde{G}}(c) \int_{0}^{c} 
                    S_c(c,\ldots,c, y, 0,\ldots,0, 0)
                    h_{t}^{\partial k}(c,\ldots,c,y,0,\dots,0, 0) dy dc.
\end{align}
For comparisons on the $(K+1)$-th endpoint, a decisive win or loss requires both participants to have observed data, which occurs with probability $\xi^2$. With ties on the first $K$ endpoints, the probability of winning at this ordinal endpoint evaluates the control participant at exactly category $q$ and the treatment participant greater than $q$ across $q\in\{1,2,\ldots,Q-1\}$:
\begin{align}
    \pi_{Y_{K+1}}^t 
        &= \mathbb{P}\left(\mathcal{O}_{K+1}^t \succ \mathcal{O}_{K+1}^c, 
            \{\mathcal{O}_i^t \simeq \mathcal{O}_i^c\}_{i=1}^{K}, R_{K+1}^t=1, R_{K+1}^c=1\right) \notag\\ 
        &= 2 \xi^2 \int_{0}^{\infty} \widetilde{g}(c) S_{\widetilde{G}}(c) \sum_{q=1}^{Q-1} \mathbb{P}\left(\min(\{Y_i^t\}_{i=1}^K) > c, Y_{K+1}^t > q\right) \mathbb{P}\left(\min(\{Y_i^c\}_{i=1}^K) > c, Y_{K+1}^c = q\right) dc \notag\\
        &= 2 \xi^2 \int_{0}^{\infty} \widetilde{g}(c)   S_{\widetilde{G}}(c) \sum_{q=1}^{Q-1} 
                    S_t(c,\ldots,c, q) \left\{ S_c(c,\ldots,c, q-1) - S_c(c,\ldots,c, q) \right\} dc. 
\end{align}
Analogously, the probability of losing at this ordinal endpoint evaluates the treatment participant at exactly category $q$ and the control participant greater than $q$ across $q\in\{1,2,\ldots,Q-1\}$:
\begin{align}
    \pi_{Y_{K+1}}^c 
        &= \mathbb{P}\left(\mathcal{O}_{K+1}^t \prec \mathcal{O}_{K+1}^c, 
            \{\mathcal{O}_i^t \simeq \mathcal{O}_i^c\}_{i=1}^{K}, R_{K+1}^t=1, R_{K+1}^c=1\right) \notag\\ 
        &= 2 \xi^2 \int_{0}^{\infty} \widetilde{g}(c)   S_{\widetilde{G}}(c) \sum_{q=1}^{Q-1} 
                    S_c(c,\ldots,c, q) \left\{ S_t(c,\ldots,c, q-1) - S_t(c,\ldots,c, q) \right\} dc. 
\end{align}
The probability of tying is then obtained as:
\begin{align}
    \mathbb{P}\left(\bm{\mathcal{O}}^t \simeq \bm{\mathcal{O}}^c\right)
        &= 2\int_{0}^{\infty} \widetilde{g}(c)   S_{\widetilde{G}}(c) \Bigg[ (1 - \xi^2) S_t(c,\ldots,c, 0) S_c(c,\ldots,c, 0) \notag\\
            &\quad + \xi^2 \sum_{q=1}^{Q} 
                    \left\{ S_t(c,\ldots,c, q-1) - S_t(c,\ldots,c, q) \right\} 
                    \left\{ S_c(c,\ldots,c, q-1) - S_c(c,\ldots,c, q) \right\} \Bigg] dc. 
\end{align}

\section{
{Additional Numerical Studies for Evaluation of the Proposed Procedure}
}
\label{apx:SIM1}
This section presents complete numerical results for Section 3.1. Table~\ref{tab:apx_SIM1} summarizes the comparison between the  simulation-based empirical results and the corresponding estimates obtained from the formula-based approach under various design configurations as specified in Section 3.1. Consistent with the findings reported in the main text, the agreement between the two approaches remains strong across all additional scenarios.

\begin{longtable}{@{}lrrrrrr@{}}
    \caption{Estimated $\text{WR}$ and $p_{\text{tie}}$ obtained from formulas ($\widetilde{\text{WR}},\widetilde {p_{\text{tie}}}$) and simulations 
      ($\widehat{\text{WR}},\hat{p_{\text{tie}}}$) across treatment effect magnitudes ($\bm \alpha$),
      follow-up times ($s$), and correlation levels (in Kendall's $\tau$).}
    \label{tab:apx_SIM1}\\

    \toprule
    $\bm \alpha$ & \multicolumn{1}{l}{$s$} & \multicolumn{1}{l}{$\tau$} & \multicolumn{1}{l}{$\widetilde{\text{WR}}$} & \multicolumn{1}{l}{$\widehat{\text{WR}}$} & \multicolumn{1}{l}{$\widetilde p_{\text{tie}}$(\%)} & \multicolumn{1}{l}{$\widehat{p_{\text{tie}}}$(\%)} \\
    \midrule
    \endfirsthead

    \caption[]{Estimated $\text{WR}$ and $p_{\text{tie}}$ (continued).}\\
    \toprule
    $\bm \alpha$ & \multicolumn{1}{l}{$s$} & \multicolumn{1}{l}{$\tau$} & \multicolumn{1}{l}{$\widetilde{\text{WR}}$} & \multicolumn{1}{l}{$\widehat{\text{WR}}$} & \multicolumn{1}{l}{$\widetilde p_{\text{tie}}$(\%)} & \multicolumn{1}{l}{$\widehat{p_{\text{tie}}}$(\%)} \\
    \midrule
    \endhead

    \midrule
    \multicolumn{7}{r}{\textit{Continued on next page}}\\
    \endfoot
    
    \bottomrule
    \endlastfoot
    
    $(0,0,0)$    & 250   & 0     & 1.00  & 1.01  & 44.00 & 43.62 \\
    $(0,0,0)$    & 250   & 0.3   & 1.00  & 1.01  & 53.58 & 53.13 \\
    $(0,0,0)$    & 250   & 0.5   & 1.00  & 1.01  & 59.10 & 58.71 \\
    $(0,0,0)$    & 250   & 0.8   & 1.00  & 1.00  & 65.43 & 65.09 \\
    $(0,0,0)$    & 500   & 0     & 1.00  & 1.01  & 9.13  & 9.12 \\
    $(0,0,0)$    & 500   & 0.3   & 1.00  & 1.00  & 15.59 & 15.46 \\
    $(0,0,0)$    & 500   & 0.5   & 1.00  & 1.00  & 20.65 & 20.53 \\
    $(0,0,0)$    & 500   & 0.8   & 1.00  & 1.00  & 27.83 & 27.72 \\
    $(0,0,0)$    & 750   & 0     & 1.00  & 1.01  & 4.45  & 4.49 \\
    $(0,0,0)$    & 750   & 0.3   & 1.00  & 1.00  & 7.27  & 7.23 \\
    $(0,0,0)$    & 750   & 0.5   & 1.00  & 1.00  & 9.97  & 9.93 \\
    $(0,0,0)$    & 750   & 0.8   & 1.00  & 1.00  & 14.52 & 14.47 \\
    $(0,0,0)$    & 1000  & 0     & 1.00  & 1.01  & 3.83  & 3.86 \\
    $(0,0,0)$    & 1000  & 0.3   & 1.00  & 1.00  & 5.45  & 5.42 \\
    $(0,0,0)$    & 1000  & 0.5   & 1.00  & 1.00  & 7.00  & 6.98 \\
    $(0,0,0)$    & 1000  & 0.8   & 1.00  & 1.00  & 9.81  & 9.79 \\
    $(0,0,0)$    & 1250  & 0     & 1.00  & 1.00  & 3.74  & 3.77 \\
    $(0,0,0)$    & 1250  & 0.3   & 1.00  & 1.00  & 5.05  & 5.02 \\
    $(0,0,0)$    & 1250  & 0.5   & 1.00  & 1.00  & 6.17  & 6.17 \\
    $(0,0,0)$    & 1250  & 0.8   & 1.00  & 1.00  & 8.14  & 8.12 \\
    $(0,0,0)$    & 1500  & 0     & 1.00  & 1.00  & 3.73  & 3.76 \\
    $(0,0,0)$    & 1500  & 0.3   & 1.00  & 1.00  & 4.96  & 4.94 \\
    $(0,0,0)$    & 1500  & 0.5   & 1.00  & 1.00  & 5.94  & 5.94 \\
    $(0,0,0)$    & 1500  & 0.8   & 1.00  & 1.00  & 7.55  & 7.53 \\
    $(0.2, 0.3, 0.1)$    & 250   & 0     & 1.24  & 1.25  & 47.22 & 46.75 \\
    $(0.2, 0.3, 0.1)$    & 250   & 0.3   & 1.24  & 1.25  & 56.61 & 56.22 \\
    $(0.2, 0.3, 0.1)$    & 250   & 0.5   & 1.25  & 1.25  & 61.96 & 61.54 \\
    $(0.2, 0.3, 0.1)$    & 250   & 0.8   & 1.26  & 1.27  & 68.13 & 67.78 \\
    $(0.2, 0.3, 0.1)$    & 500   & 0     & 1.26  & 1.27  & 11.00 & 10.92 \\
    $(0.2, 0.3, 0.1)$    & 500   & 0.3   & 1.26  & 1.26  & 18.24 & 18.09 \\
    $(0.2, 0.3, 0.1)$    & 500   & 0.5   & 1.26  & 1.27  & 23.70 & 23.51 \\
    $(0.2, 0.3, 0.1)$    & 500   & 0.8   & 1.28  & 1.29  & 31.38 & 31.22 \\
    $(0.2, 0.3, 0.1)$    & 750   & 0     & 1.27  & 1.27  & 5.20  & 5.17 \\
    $(0.2, 0.3, 0.1)$    & 750   & 0.3   & 1.27  & 1.27  & 8.63  & 8.58 \\
    $(0.2, 0.3, 0.1)$    & 750   & 0.5   & 1.27  & 1.28  & 11.80 & 11.74 \\
    $(0.2, 0.3, 0.1)$    & 750   & 0.8   & 1.29  & 1.29  & 17.07 & 17.01 \\
    $(0.2, 0.3, 0.1)$    & 1000  & 0     & 1.26  & 1.27  & 4.27  & 4.25 \\
    $(0.2, 0.3, 0.1)$    & 1000  & 0.3   & 1.27  & 1.27  & 6.22  & 6.20 \\
    $(0.2, 0.3, 0.1)$    & 1000  & 0.5   & 1.27  & 1.28  & 8.10  & 8.07 \\
    $(0.2, 0.3, 0.1)$    & 1000  & 0.8   & 1.29  & 1.29  & 11.50 & 11.47 \\
    $(0.2, 0.3, 0.1)$    & 1250  & 0     & 1.26  & 1.26  & 4.12  & 4.10 \\
    $(0.2, 0.3, 0.1)$    & 1250  & 0.3   & 1.26  & 1.26  & 5.62  & 5.61 \\
    $(0.2, 0.3, 0.1)$    & 1250  & 0.5   & 1.27  & 1.27  & 6.94  & 6.92 \\
    $(0.2, 0.3, 0.1)$    & 1250  & 0.8   & 1.28  & 1.29  & 9.33  & 9.32 \\
    $(0.2, 0.3, 0.1)$    & 1500  & 0     & 1.25  & 1.26  & 4.10  & 4.08 \\
    $(0.2, 0.3, 0.1)$    & 1500  & 0.3   & 1.26  & 1.26  & 5.47  & 5.46 \\
    $(0.2, 0.3, 0.1)$    & 1500  & 0.5   & 1.26  & 1.27  & 6.59  & 6.57 \\
    $(0.2, 0.3, 0.1)$    & 1500  & 0.8   & 1.28  & 1.28  & 8.49  & 8.47 \\
    $(0.1, 0.2, 0.3)$    & 250   & 0     & 1.23  & 1.23  & 47.48 & 46.95 \\
    $(0.1, 0.2, 0.3)$    & 250   & 0.3   & 1.24  & 1.25  & 56.90 & 56.44 \\
    $(0.1, 0.2, 0.3)$    & 250   & 0.5   & 1.24  & 1.26  & 62.24 & 61.83 \\
    $(0.1, 0.2, 0.3)$    & 250   & 0.8   & 1.24  & 1.25  & 68.18 & 67.81 \\
    $(0.1, 0.2, 0.3)$    & 500   & 0     & 1.20  & 1.20  & 11.17 & 11.10 \\
    $(0.1, 0.2, 0.3)$    & 500   & 0.3   & 1.21  & 1.21  & 18.50 & 18.34 \\
    $(0.1, 0.2, 0.3)$    & 500   & 0.5   & 1.22  & 1.22  & 24.01 & 23.85 \\
    $(0.1, 0.2, 0.3)$    & 500   & 0.8   & 1.22  & 1.22  & 31.44 & 31.28 \\
    $(0.1, 0.2, 0.3)$    & 750   & 0     & 1.17  & 1.17  & 5.27  & 5.25 \\
    $(0.1, 0.2, 0.3)$    & 750   & 0.3   & 1.18  & 1.19  & 8.77  & 8.70 \\
    $(0.1, 0.2, 0.3)$    & 750   & 0.5   & 1.19  & 1.20  & 12.00 & 11.94 \\
    $(0.1, 0.2, 0.3)$    & 750   & 0.8   & 1.20  & 1.21  & 17.12 & 17.04 \\
    $(0.1, 0.2, 0.3)$    & 1000  & 0     & 1.16  & 1.16  & 4.31  & 4.30 \\
    $(0.1, 0.2, 0.3)$    & 1000  & 0.3   & 1.17  & 1.17  & 6.30  & 6.25 \\
    $(0.1, 0.2, 0.3)$    & 1000  & 0.5   & 1.18  & 1.18  & 8.22  & 8.17 \\
    $(0.1, 0.2, 0.3)$    & 1000  & 0.8   & 1.19  & 1.19  & 11.53 & 11.50 \\
    $(0.1, 0.2, 0.3)$    & 1250  & 0     & 1.15  & 1.15  & 4.16  & 4.14 \\
    $(0.1, 0.2, 0.3)$    & 1250  & 0.3   & 1.16  & 1.16  & 5.68  & 5.63 \\
    $(0.1, 0.2, 0.3)$    & 1250  & 0.5   & 1.16  & 1.17  & 7.03  & 6.99 \\
    $(0.1, 0.2, 0.3)$    & 1250  & 0.8   & 1.18  & 1.18  & 9.35  & 9.33 \\
    $(0.1, 0.2, 0.3)$    & 1500  & 0     & 1.14  & 1.14  & 4.13  & 4.12 \\
    $(0.1, 0.2, 0.3)$    & 1500  & 0.3   & 1.15  & 1.16  & 5.52  & 5.48 \\
    $(0.1, 0.2, 0.3)$    & 1500  & 0.5   & 1.16  & 1.16  & 6.65  & 6.62 \\
    $(0.1, 0.2, 0.3)$    & 1500  & 0.8   & 1.17  & 1.17  & 8.51  & 8.49 \\
    $(0.2, 0.2, 0.2)$    & 250   & 0     & 1.22  & 1.23  & 47.16 & 46.66 \\
    $(0.2, 0.2, 0.2)$    & 250   & 0.3   & 1.22  & 1.23  & 56.54 & 56.10 \\
    $(0.2, 0.2, 0.2)$    & 250   & 0.5   & 1.22  & 1.23  & 61.86 & 61.43 \\
    $(0.2, 0.2, 0.2)$    & 250   & 0.8   & 1.22  & 1.23  & 67.91 & 67.54 \\
    $(0.2, 0.2, 0.2)$    & 500   & 0     & 1.22  & 1.22  & 10.97 & 10.88 \\
    $(0.2, 0.2, 0.2)$    & 500   & 0.3   & 1.22  & 1.23  & 18.17 & 18.00 \\
    $(0.2, 0.2, 0.2)$    & 500   & 0.5   & 1.22  & 1.23  & 23.59 & 23.45 \\
    $(0.2, 0.2, 0.2)$    & 500   & 0.8   & 1.22  & 1.23  & 31.08 & 30.89 \\
    $(0.2, 0.2, 0.2)$    & 750   & 0     & 1.22  & 1.22  & 5.19  & 5.17 \\
    $(0.2, 0.2, 0.2)$    & 750   & 0.3   & 1.22  & 1.23  & 8.59  & 8.51 \\
    $(0.2, 0.2, 0.2)$    & 750   & 0.5   & 1.22  & 1.23  & 11.73 & 11.67 \\
    $(0.2, 0.2, 0.2)$    & 750   & 0.8   & 1.22  & 1.23  & 16.85 & 16.76 \\
    $(0.2, 0.2, 0.2)$    & 1000  & 0     & 1.22  & 1.22  & 4.26  & 4.26 \\
    $(0.2, 0.2, 0.2)$    & 1000  & 0.3   & 1.22  & 1.23  & 6.20  & 6.15 \\
    $(0.2, 0.2, 0.2)$    & 1000  & 0.5   & 1.22  & 1.23  & 8.05  & 8.02 \\
    $(0.2, 0.2, 0.2)$    & 1000  & 0.8   & 1.22  & 1.23  & 11.35 & 11.31 \\
    $(0.2, 0.2, 0.2)$    & 1250  & 0     & 1.22  & 1.22  & 4.12  & 4.11 \\
    $(0.2, 0.2, 0.2)$    & 1250  & 0.3   & 1.22  & 1.23  & 5.60  & 5.56 \\
    $(0.2, 0.2, 0.2)$    & 1250  & 0.5   & 1.22  & 1.23  & 6.91  & 6.88 \\
    $(0.2, 0.2, 0.2)$    & 1250  & 0.8   & 1.22  & 1.23  & 9.22  & 9.20 \\
    $(0.2, 0.2, 0.2)$    & 1500  & 0     & 1.22  & 1.22  & 4.09  & 4.09 \\
    $(0.2, 0.2, 0.2)$    & 1500  & 0.3   & 1.22  & 1.23  & 5.45  & 5.42 \\
    $(0.2, 0.2, 0.2)$    & 1500  & 0.5   & 1.22  & 1.23  & 6.56  & 6.53 \\
    $(0.2, 0.2, 0.2)$    & 1500  & 0.8   & 1.22  & 1.23  & 8.40  & 8.38 \\
    
\end{longtable}

\newpage

Table~\ref{tab:apx_SIM1_TYPEI} reports the formula-based and simulation-based empirical type I error rates under the null scenario. The formula-based procedure correctly preserves the nominal 5\% level under the null. The empirical type I error rates are within the Monte Carlo variation range for the nominal level in nearly all scenarios considered, with one setting showing slight conservatism (3.75\%). Overall, no significant type I error inflation is observed in the included simulation scenarios.

\begin{table}[htbp]
  \centering
  \caption{Formula-based and simulation-based empirical type I error rates under the null scenario across maximal follow-up times ($s$) and correlation levels (in Kendall's $\tau$). The acceptable range of empirical type I error rate with Monte Carlo variation under 2000 replicates is 4.04\% to 5.96\% for the 5\% nominal level.}
    \begin{tabular}{lrrrr}
    \toprule
    $\bm \alpha$ & \multicolumn{1}{l}{$s$} & \multicolumn{1}{l}{$\tau$} & \multicolumn{1}{l}{Formula-based Type I Error} & \multicolumn{1}{l}{Empirical Type I Error} \\
    \midrule
    $(0,0,0)$ & 250   & 0     & 5.00  & 5.80 \\
    $(0,0,0)$ & 250   & 0.3   & 5.00  & 4.65 \\
    $(0,0,0)$ & 250   & 0.5   & 5.00  & 4.40 \\
    $(0,0,0)$ & 250   & 0.8   & 5.00  & 4.95 \\
    $(0,0,0)$ & 500   & 0     & 5.00  & 5.00 \\
    $(0,0,0)$ & 500   & 0.3   & 5.00  & 5.15 \\
    $(0,0,0)$ & 500   & 0.5   & 5.00  & 4.70 \\
    $(0,0,0)$ & 500   & 0.8   & 5.00  & 3.75 \\
    $(0,0,0)$ & 750   & 0     & 5.00  & 5.20 \\
    $(0,0,0)$ & 750   & 0.3   & 5.00  & 5.20 \\
    $(0,0,0)$ & 750   & 0.5   & 5.00  & 4.15 \\
    $(0,0,0)$ & 750   & 0.8   & 5.00  & 4.40 \\
    $(0,0,0)$ & 1000  & 0     & 5.00  & 5.35 \\
    $(0,0,0)$ & 1000  & 0.3   & 5.00  & 5.80 \\
    $(0,0,0)$ & 1000  & 0.5   & 5.00  & 4.60 \\
    $(0,0,0)$ & 1000  & 0.8   & 5.00  & 4.10 \\
    $(0,0,0)$ & 1250  & 0     & 5.00  & 5.50 \\
    $(0,0,0)$ & 1250  & 0.3   & 5.00  & 5.65 \\
    $(0,0,0)$ & 1250  & 0.5   & 5.00  & 4.95 \\
    $(0,0,0)$ & 1250  & 0.8   & 5.00  & 4.15 \\
    $(0,0,0)$ & 1500  & 0     & 5.00  & 4.90 \\
    $(0,0,0)$ & 1500  & 0.3   & 5.00  & 5.25 \\
    $(0,0,0)$ & 1500  & 0.5   & 5.00  & 4.80 \\
    $(0,0,0)$ & 1500  & 0.8   & 5.00  & 4.05 \\
    \bottomrule
    \end{tabular}%
  \label{tab:apx_SIM1_TYPEI}%
\end{table}%

\FloatBarrier
\section{
{
Additional Numerical Studies on the Influence of Correlation for Study Planning
}
}
\label{apx:SIM2}

\subsection{Simulation-based Sample Size Determination for  Scenarios with a Fixed Study Length}
\label{apx:SIM2_simulated}

In this subsection, we examine the impact of correlation using simulation-based sample size approximations for scenarios with a fixed study length as presented in Section 3.2. 

To estimate the required sample size by simulation, we evaluate a range of candidate sample sizes. For each candidate, we generate 2,000 datasets and estimate empirical power as the proportion rejecting the null hypothesis at a two-sided significance level of 0.05. For the target power of 80\% ($\beta=0.2$),  a candidate sample size is considered acceptable if its empirical power is between 79\% and 81\%.
 
\begin{figure}
\centerline{\includegraphics[width=6in]{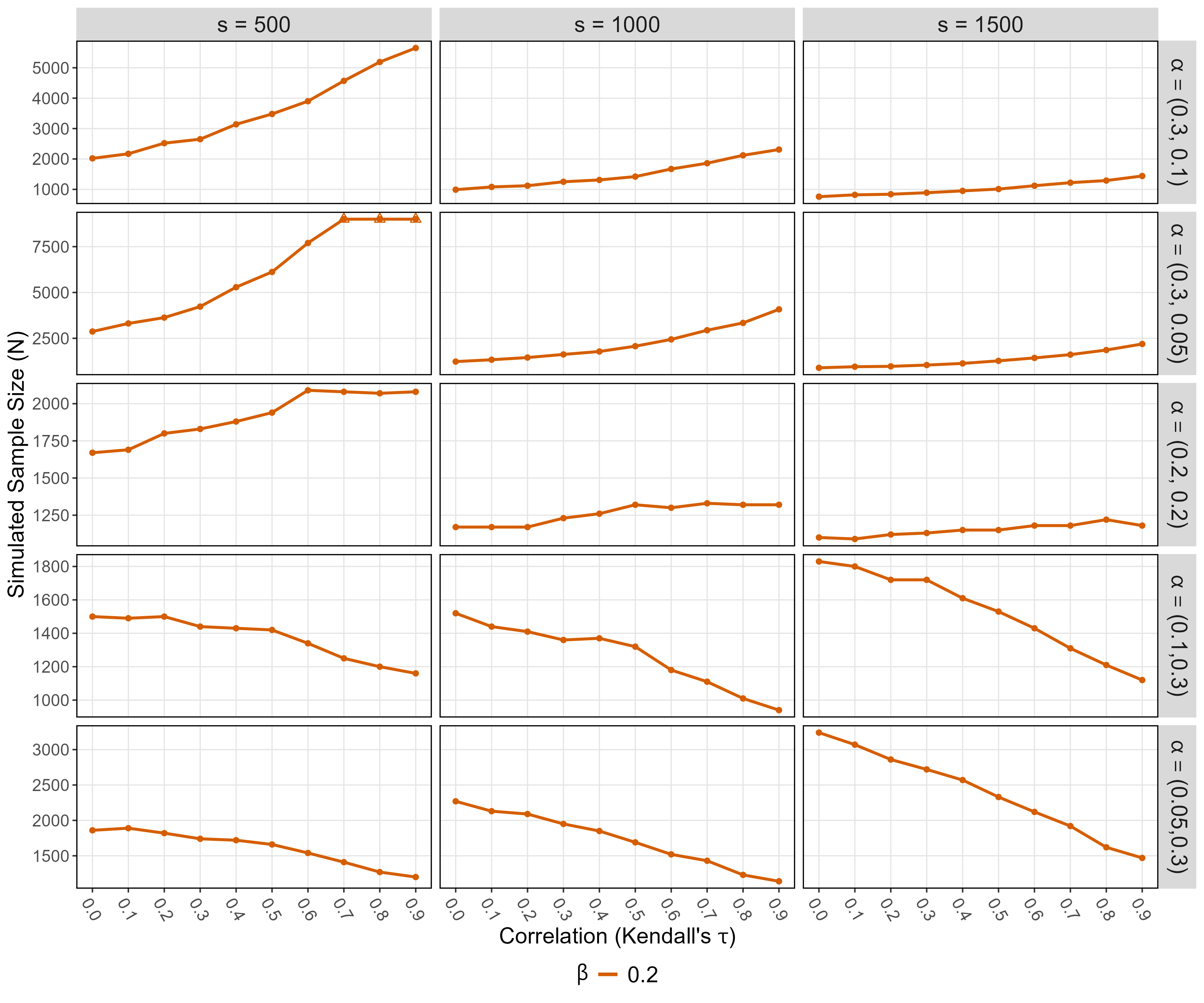}}
\caption{Simulation-based sample size estimation across treatment effect magnitudes ($\bm \alpha$), correlation levels ($\tau$), and follow-up times ($s$), targeting type II error rate $\beta=0.2$. \label{fig:apx_SIM2_simulated_SS}}
\end{figure}

 The results are shown in Figure~\ref{fig:apx_SIM2_simulated_SS}. 
 The simulation-based results show trends consistent with those from the formula-based approach. In particular, the direction and magnitude of changes in the required sample size across correlation levels and treatment-effect settings are similar. Across all scenarios, the formula-based sample sizes are comparable to or larger than the simulation-based estimates, further demonstrating the conservative nature of the YG formula under the settings considered. 
 The simulation-based curves are slightly less smooth because of Monte Carlo variation and the 1\% tolerance used to select acceptable sample sizes.

\FloatBarrier
\subsection{Numerical Studies for Scenarios with a Fixed Follow-up Length }

In this subsection, we assess the impact of correlation on sample size requirements under a fixed follow-up time and examine how this impact varies across dropout rates. 

Specifically, each participant is assumed to have a scheduled follow-up time of 1000 days, regardless of their entry time. Independent dropout censoring follows an exponential distribution with hazard rate (per day) $\lambda_d \in \{0, 0.00015, 0.00035, 0.00065\}$, representing no, low, medium, and high dropout censoring, respectively. These settings correspond to expected dropout censoring rates of 0\%, 13.9\%, 29.5\%, and 47.8\% by 1000 days. Unless otherwise noted, all other design features remain the same as those in Section 3.2, including the baseline hazard rates and the five treatment effect magnitude configurations.

\begin{figure}
\centerline{\includegraphics[width=6.5in]{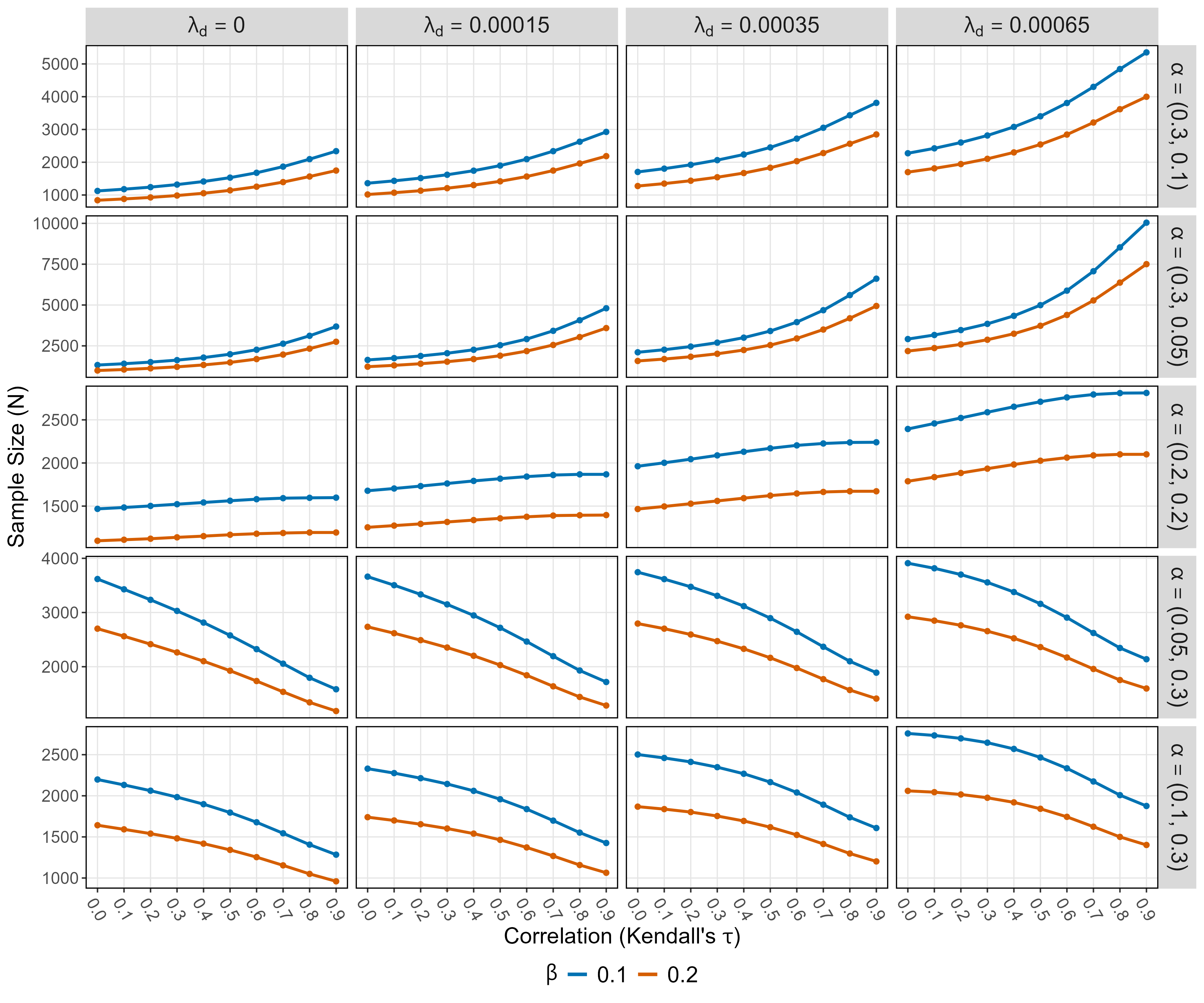}}
\caption{
Formula-based sample size estimation across treatment effect magnitudes ($\bm \alpha$), correlation levels ($\tau$), dropout censoring hazard rate ($\lambda_d$), and targeting type II error rates ($\beta$)} \label{fig:apx_SIM3_SS}
\end{figure}

\begin{sidewaysfigure}
\centerline{\includegraphics[width=\textwidth]{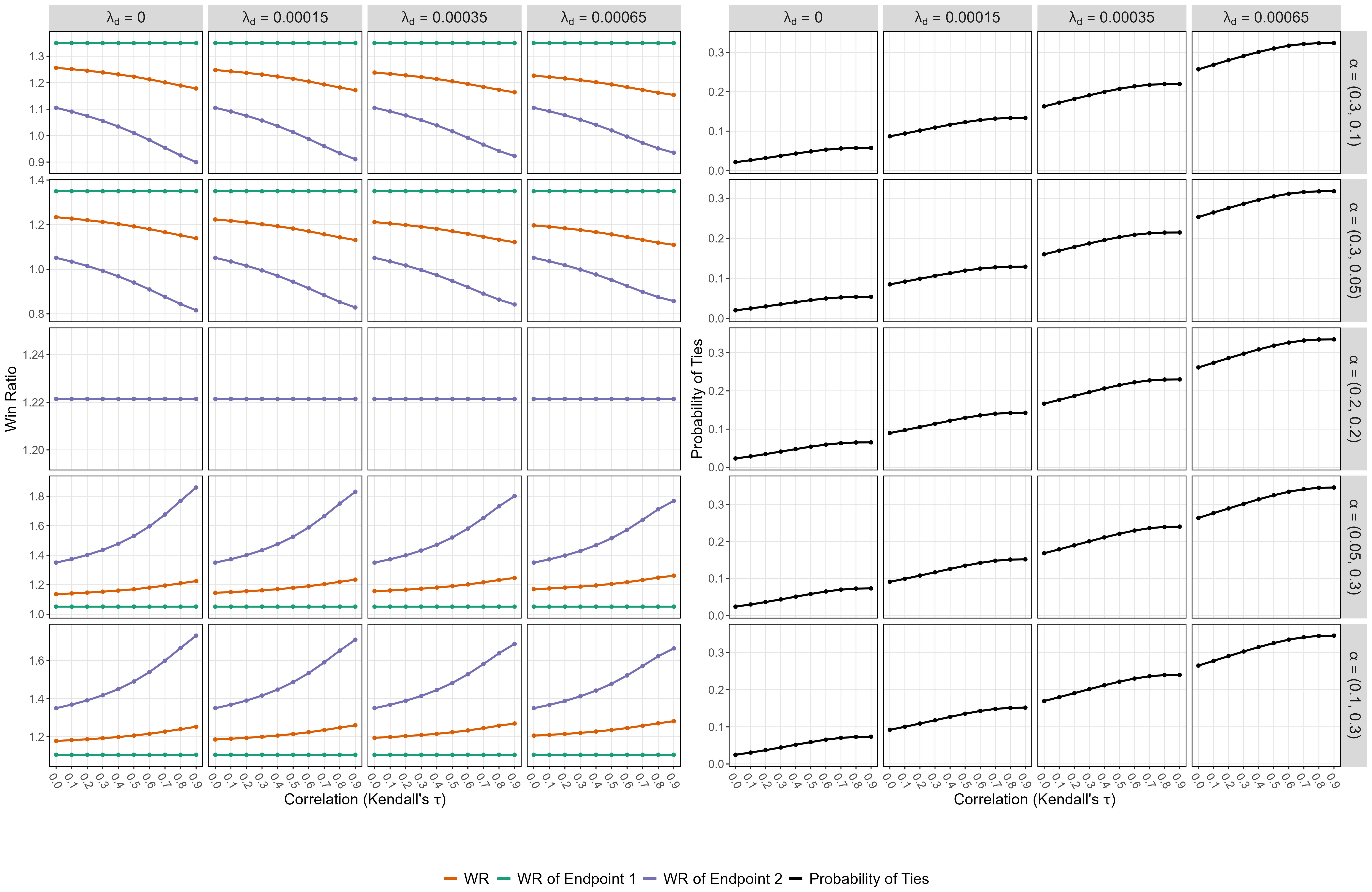}}
\caption{Win ratio (WR) and probability of ties under the fixed follow-up time setup (1000 days).
\textit{Left panel:} overall and endpoint-specific WR.
\textit{Right panel:} probability of ties ($p_{\text{tie}}$).
Both are shown across treatment effect magnitudes ($\bm{\alpha}$),
correlation levels ($\tau$), and dropout censoring hazard rates ($\lambda_d$).
\label{fig:apx_SIM3_WR_ptie_trend}}
\end{sidewaysfigure}

\begin{figure}
\centerline{\includegraphics[width=\textwidth]{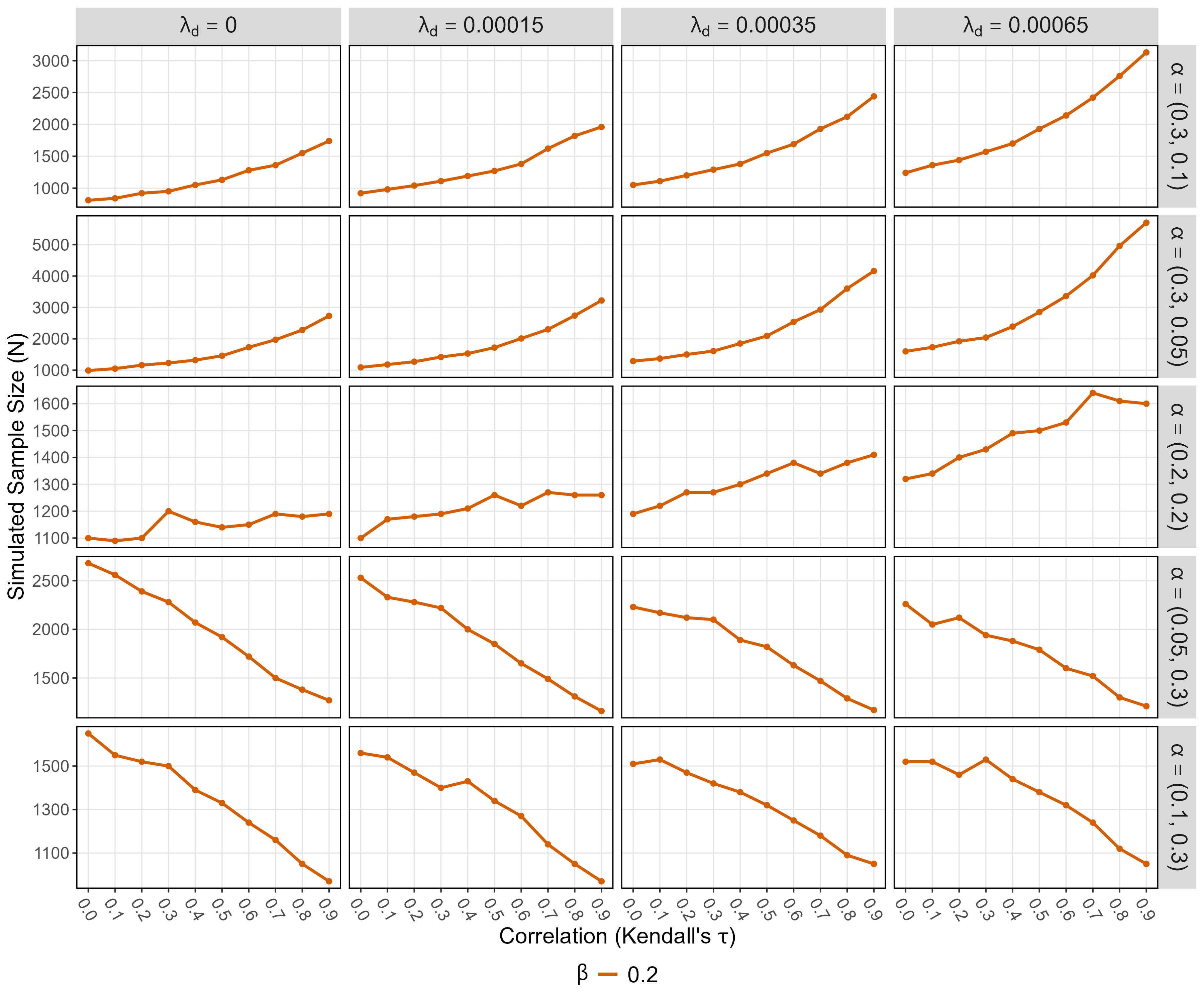}}
\caption{Simulation-based sample size estimation across treatment effect magnitudes ($\bm \alpha$), correlation levels ($\tau$), and dropout censoring hazard rate ($\lambda_d$), targeting type II error rate $\beta=0.2$. \label{fig:apx_SIM3_simulated_SS}}
\end{figure}

The results are presented in Figure~\ref{fig:apx_SIM3_SS} and Figure~\ref{fig:apx_SIM3_WR_ptie_trend}. The formula-based results under the fixed follow-up time setup show patterns that are generally consistent with those observed under the fixed study-length setup in Section 3.2. In particular, when the treatment effect is stronger on the first endpoint, increasing correlation tends to increase the required sample size, whereas when the treatment effect is stronger on the second endpoint, higher correlation generally reduces the required sample size. This pattern again reflects the relative contribution of the correlated endpoints to the overall WR, depending on where the treatment effect is concentrated. In addition, under the fixed follow-up time setup, higher dropout censoring leads to a larger required sample size. This is because heavier dropout censoring results in more unresolved pairwise comparisons and thus a higher $p_{\text{tie}}$, as also reflected in the right panel of Figure~\ref{fig:apx_SIM3_WR_ptie_trend}.
We also present the corresponding simulation-based sample size approximations in Figure~\ref{fig:apx_SIM3_simulated_SS}, using the same approach as described in Section~\ref{apx:SIM2_simulated} and targeting $\beta=0.2$. The trend in the required sample size across correlation levels is generally consistent with the formula-based results, although the simulation-based curves are less smooth because of Monte Carlo variation and the tolerance range used for the target power.

\FloatBarrier
\section{{Additional Case Studies}}

\subsection{Additional descriptive results of Case Studies} \label{sec:apx_case_detail}

This subsection reports additional descriptive results from the WR analysis based on the observed trial data. Specifically, for SPRINT and STICH case studies, we summarize the endpoint-level proportions of wins, losses, and ties arising from the pairwise comparisons in Table~\ref{tab:apx_SPRINT_decomp} and Table~\ref{tab:apx_STICH_decomp}, respectively.

\begin{table}[htbp]
  \centering
  \caption{Endpoint-level proportions of wins, ties, and losses for two endpoints in WR analysis of the observed SPRINT data.}
    \begin{tabular}{p{16em}rrrr}
    \toprule
    Endpoint & \multicolumn{1}{l}{Win(\%)} & \multicolumn{1}{l}{Tie(\%)} & \multicolumn{1}{l}{Loss(\%)} & \multicolumn{1}{l}{Endpoint-level WR} \\
    \midrule
    Time to cardiovascular-related death & 3.64  & 93.69 & 2.67  & 1.36 \\
    Time to first non-fatal cardiovascular event & 4.24  & 86.01 & 3.44  & 1.23 \\
    \bottomrule
    \end{tabular}%
  \label{tab:apx_SPRINT_decomp}%
\end{table}%

\begin{table}[htbp]
  \centering
  \caption{Endpoint-level proportions of wins, ties, and losses for two endpoints in WR analysis of the observed STICH data under two-endpoint setup}
    \begin{tabular}{p{16em}rrrr}
    \toprule
    \multicolumn{1}{l}{Endpoint} & \multicolumn{1}{l}{Win(\%)} & \multicolumn{1}{l}{Tie(\%)} & \multicolumn{1}{l}{Loss(\%)} & \multicolumn{1}{l}{Endpoint-level WR} \\
    \midrule
    Time to cardiovascular-related death & 24.67 & 54.27 & 21.06 & 1.17 \\
    \multicolumn{1}{p{16em}}{Time to first cardiovascular-related hospitalization} & 20.23 & 21.38 & 12.66 & 1.60\\
    \bottomrule
    \end{tabular}%
  \label{tab:apx_STICH_decomp}%
\end{table}%

\subsection{Three-component Composite Endpoint Setup with the STICH Trial}
In this three-endpoint setup with the STICH trial, we consider a hierarchical endpoint with three layers: (1) time to cardiovascular-related death, (2) time to first cardiovascular-related hospitalization, followed by (3) time to first revascularization.
The endpoint-level proportions of wins, losses, and ties for the included endpoints in the observed STICH data are presented in Table~\ref{tab:apx_STICH_decomp_three}.
In STICH, the accrual period was approximately 1,746 days, and the maximum follow-up time was 3,051 days. The estimated hazard rate of dropout (including non-cardiovascular death) was $3.88 \times 10^{-5}$ per day. The estimated hazard rates for the included three endpoints in the treatment group were $1.92 \times 10^{-4}$, $3.22 \times 10^{-4}$, and $3.06 \times 10^{-5}$ per day, respectively, while the corresponding hazard rates in the control group were $2.38 \times 10^{-4}$, $5.11 \times 10^{-4}$, and $1.84 \times 10^{-4}$ per day. The pairwise associations among the three endpoints are first estimated using Kendall’s $\tau$ and then converted to Pearson correlations via the conventional relation $\varrho = \sin(\pi\tau/2)$ \citep{kruskal1958ordinal}, yielding the estimated correlation matrix
\[
\hat{\boldsymbol{\Sigma}} = \begin{pmatrix} 1.00 & 0.25 & 0.02 \\ 0.25 & 1.00 & 0.51 \\ 0.02 & 0.51 & 1.00 \end{pmatrix},
\]
where the rows and columns correspond to the three endpoints in their hierarchical order.
We observe that the correlation between layers 2 and 3 is the strongest ($\hat{\varrho}_{23} = 0.51$), followed by the correlation between layers 1 and 2 ($\hat{\varrho}_{12} = 0.25$), while the correlation between layers 1 and 3 ($\hat{\varrho}_{13} = 0.02$) appears to be weak. With the estimated correlation matrix, the Gaussian copula emerges as a natural candidate to construct the joint survival function, due to its capacity to accommodate heterogeneous pairwise association parameters.

\begin{table}[htbp]
  \centering
  \caption{Endpoint-level proportions of wins, ties, and losses for endpoints in WR analysis of the observed STICH data under three-endpoint setup}
    \begin{tabular}{p{16em}rrrr}
    \toprule
    \multicolumn{1}{l}{Endpoint} & \multicolumn{1}{l}{Win(\%)} & \multicolumn{1}{l}{Tie(\%)} & \multicolumn{1}{l}{Loss(\%)} & \multicolumn{1}{l}{Endpoint-level WR} \\
    \midrule
    Time to cardiovascular-related death & 24.67 & 54.27 & 21.06 & 1.17 \\
    \multicolumn{1}{p{16em}}{Time to first cardiovascular-related hospitalization} & 20.23 & 21.38 & 12.66 & 1.60 \\
    Time to first revascularization & 1.74  & 19.50 & 0.14  & 12.81 \\
    \bottomrule
    \end{tabular}%
  \label{tab:apx_STICH_decomp_three}%
\end{table}%

To construct a sample size table across different correlation levels, we characterize a family of correlation matrices by defining $\tilde{\boldsymbol{\Sigma}} = \gamma \hat{\boldsymbol{\Sigma}} + (1 - \gamma) \mathbf{I}$, where $\mathbf{I}$ is the identity matrix of the same dimension as $\hat{\boldsymbol{\Sigma}}$ and $\gamma \in [0, \bar{\gamma}]$ is a scaling coefficient that controls the overall correlation level. 
We select $\bar{\gamma}$ such that the largest absolute off-diagonal element in $\tilde{\boldsymbol{\Sigma}}$ reaches approximately 0.9, representing an overall high level of correlation. In our case, this corresponds to $\bar{\gamma} = 1.78$. In this study, in addition to $\gamma = 1$ (which recovers $\hat{\boldsymbol{\Sigma}}$), we consider ten additional correlation matrices with $\gamma_i$ linearly spaced over the range $[0, 1.78]$, where $\gamma = 0$ corresponds to the independence scenario.

\begin{table}[htbp]
  \centering
  \caption{ Formula-based sample size estimates for the STICH trial with three time-to-event endpoints under varying correlation levels ($\gamma$) and type II error rates ($\beta$) with the Gaussian copula. The estimates corresponding to the independence assumption ($\gamma = 0$) and to the empirically estimated correlation level (i.e., $\hat{\boldsymbol{\Sigma}}$, where $\gamma = 1$) are highlighted in bold.}
    \begin{tabular}{rrrrrrrrrrrr}
    \toprule
          & \multicolumn{11}{c}{$\gamma$} \\
    \multicolumn{1}{l}{$\beta$} & $\bm{0}$     & 0.20  & 0.40  & 0.59  & 0.79  & 0.99  & $\bm{1}$     & 1.19  & 1.38  & 1.58  & 1.78 \\
    \midrule
    0.1   & 552   & 572   & 592   & 614   & 638   & 664   & 666   & 696   & 734   & 782   & 832 \\
    0.2   & 412   & 428   & 442   & 458   & 476   & 496   & 498   & 520   & 548   & 584   & 622 \\
    \bottomrule
    \end{tabular}%
  \label{tab:STICH_3TTE_SS}%
\end{table}%

Table~\ref{tab:STICH_3TTE_SS} presents the sample size estimates under the three-component composite endpoint setup. Here, the required sample sizes increase monotonically with $\gamma$, that is, with the overall correlation level. At $\gamma = 1.78$, the largest pairwise correlation among endpoints (between layers 2 and 3) reaches $\tilde{\varrho}_{23} = 0.908$, with $\tilde{\varrho}_{12} = 0.445$ between layers 1 and 2 and $\tilde{\varrho}_{13} = 0.036$ between layers 1 and 3, representing a high-correlation scenario. By construction, the relative correlation strengths among endpoints at each $\gamma$ level are scaled to preserve the structure of $\hat{\boldsymbol{\Sigma}}$, so that the pairwise correlation pattern observed in the data is maintained across the range of overall correlation levels presented in the table.
In such scenarios, accounting for the between-endpoint correlation becomes essential at the design stage, and one may evaluate sample size across a plausible range of correlation values for a more robust design calculation.

\FloatBarrier

\end{document}